\documentclass[aps,prb,reprint,superscriptaddress,preprintnumbers,floatfix]{revtex4-2}

\usepackage{amsfonts,amsmath,amssymb,amsbsy}
\usepackage{caption}
\usepackage{float}
\usepackage[T1]{fontenc}
\usepackage{subcaption,graphicx}
\usepackage{dcolumn}
\usepackage{bm}

\usepackage{adjustbox}
\usepackage{booktabs}
\usepackage{tabularx}
\usepackage{color}
\usepackage{hyperref}
\usepackage{lettrine}
\usepackage{caption}
\begin{document}


\title{Biorthogonal quench dynamics of entanglement and quantum geometry in $\mathcal{PT}$-symmetric non-Hermitian systems}

\author{Hsueh-Hao Lu}
\affiliation{Department of Physics, National Tsing Hua University, Hsinchu 30013, Taiwan}
\author{Po-Yao Chang}%
\email{pychang@phys.nthu.edu.tw}
\affiliation{Department of Physics, National Tsing Hua University, Hsinchu 30013, Taiwan}
\affiliation{Yukawa Institute for Theoretical Physics, Kyoto University, Kyoto 606-8502, Japan}



\date{\today}

\begin{abstract}
We explore the quench dynamics of $\mathcal{PT}$-symmetric non-Hermitian systems by utilizing the biorthogonal formalism. 
We analyze quench dynamics of observable quantities, the quantum geometric tensor, and various entanglement quantities, including the entanglement entropy, the SVD entropy, and the Tu-Tzeng-Chang entropy. 
Our results show that a sudden quench into a PT-broken phase generally leads to exponential growth in these quantities, driven by the biorthogonal density matrix's non-positivity. 
In contrast to generic interacting systems, we observe a surprising linear decay in the TTC entropy for non-interacting fermionic systems.
This finding originates from the approximate spectral symmetry of the biorthogonal reduced density matrix, and we confirm our findings using the Yang-Lee and non-Hermitian XXZ models.
\end{abstract}

\maketitle

\section{\label{sec:level1}Introduction}

The study of non-Hermitian quantum systems has attracted significant attention due to unique features, such as the non-Hermitian skin effect~\cite{Yao2018, Kawabata2019, Zhang31122022} and the existence 
of exceptional points~\cite{Gong2018, Yoshida2019, Okugawa2019, Ashida2020, Bergholtz2021,Kozii2024} that have no Hermitian counterparts. In addition, these non-Hermitian quantum systems can be thought of as effective descriptions of open quantum systems and their interactions with external environments~\cite{Ashida2020,Rotter2009}.  
One interesting investigation of  non-Hermitian quantum systems is their relevance to measurement-induced entanglement transitions (MIETs)~\cite{Jian2021,PRXQuantum.2.030313,PhysRevLett.127.140601}, where entanglement dynamics are dramatically altered by quantum measurements, leading to phase transitions between volume-law and area-law entanglement scaling~\cite{Jian2021,PhysRevLett.127.140601}.

A particular realization of these MIETs occurs in the Yang-Lee edge singularity, where the singularity triggers a transition in entanglement scaling that can be effectively described using a  parity-time ($\mathcal{PT}$)-symmetric non-Hermitian Hamiltonian~\cite{Jian2021,li2022dynamicssimulationnumericalanalysis}. This connection provides a physical interpretation of how non-Hermitian interactions mimic measurement processes, leading to entanglement dynamics distinct from those in conventional Hermitian quantum systems~\cite{Rotter2009, Jian2021, Turkeshi_2023, B_csi_2021, deng2024restoringkibblezurekscalingdefect}.

A fundamental challenge in analyzing the dynamics of non-Hermitian systems lies in choosing an appropriate basis for describing their time evolution~\cite{Brody2013,Ashida2020}. 
Considering the Schr\"odinger equation of the wavefunction evolved under a non-Hermitian Hamiltonian, the density matrix constructed from the wavefunction  may not be trace-preserving.
The violation of the trace preserving property of the density matrix has a physical interpretation in open quantum systems where probability conservation can be violated.
On the other hand, the biorthogonal framework in non-Hermitian quantum systems incorporates both right and left eigenstates. Using the biorthogonal basis,
the (biorthogonal) density matrix is not Hermitian, but the trace-preserving property still holds~\cite{Chang_2020, Tu2023,Ye2024}.
This consistency in the probabilistic interpretation of quantum mechanics ensures that observable quantities and dynamics can be evaluated through the modified inner product~\cite{Brody2013, matsumoto2022embeddingyangleequantumcriticality, PhysRevLett.132.220402,Lu_2025}.



In this work, besides the observable quantities, we further investigate the quench dynamics of the quantum geometric tensor~\cite{Juan2017, Ahn2020, Holder2020} and different entanglement quantities in $\mathcal{PT}$-symmetric non-Hermitian systems using the biorthogonal formalism.  
The entanglement quantities we investigate include the entanglement entropy, the singular-value-decomposition (SVD) entropy~\cite{Parzygnat_2023}, and the generic entanglement entropy (referred to as the Tu-Tzeng-Chang (TTC) entropy)~\cite{Tu_2022}.
For the case where the system is under a sudden quench by a $\mathcal{PT}$-unbroken phase of the post-quench Hamiltonian (referred to as the $\mathcal{PT}$-unbroken case),
the dynamics of all quantities are governed by a unitary-like evolution. On the other hand, for the case where the system is under a sudden quench by a $\mathcal{PT}$-broken phase of the post-quench Hamiltonian (referred to as the $\mathcal{PT}$-broken case), the off-diagonal elements of the biorthogonal density matrix are exponential functions of time, which lead to all quantities growing exponentially in time.
Interestingly, we find that for non-interacting fermionic systems in the $\mathcal{PT}$-broken case, the TTC entropy decays linearly in time, rather than growing exponentially.
This linear-decay property comes from the approximate spectral symmetry in the biorthogonal reduced density matrix (RDM) which is broken in the interacting systems.
We demonstrate these properties using the Yang-Lee model~\cite{PhysRevB.106.174517,yan2024dissipativedynamicalphasetransition,PhysRevResearch.4.033250}
and the non-Hermitian XXZ model~\cite{Tu2023}.
Results are summarized in Table~\ref{tab:entropy_comparison}.

\begin{table*}[htbp]
\centering
  \renewcommand{\arraystretch}{1.4}  
  \begin{tabularx}{\textwidth}{@{} >{\raggedright\arraybackslash}X 
                                   >{\raggedright\arraybackslash}X 
                                   >{\raggedright\arraybackslash}X @{}}
    \toprule
    \textbf{Quantities} & \textbf{Transition Point} & \textbf{Long-time Dynamics} \\
    \midrule
    Observables $\langle \hat{O} \rangle $
     & $\mathcal{PT}$-transition 
     & $\mathcal{PT}$-unbroken case: oscillation.
        $\mathcal{PT}$-broken case: exponential growth.\\

    Quantum metric tensor $\chi^{RL}_{\alpha\beta} $
     & $\mathcal{PT}$-transition 
     & $\mathcal{PT}$-unbroken case: oscillation.
         $\mathcal{PT}$-broken case: exponential growth.\\

    Entanglement entropy $S_A$ 
      & $\mathcal{PT}$-transition and Level crossings
      & $\mathcal{PT}$-unbroken case: oscillation.
          $\mathcal{PT}$-broken case: saturate to a constant  and has a volume to area law transition across the level crossing point~\cite{Jian2021}. \\
    \addlinespace[4pt]

    SVD  entropy $S\bigl(\rho_A^{1|2}\bigr)$~\cite{Parzygnat_2023}
      & $\mathcal{PT}$-transition and Level crossings 
      & $\mathcal{PT}$-unbroken case: oscillation.
          $\mathcal{PT}$-broken case: saturate to a constant  and has a volume to area law transition across the level crossing point. \\

    TTC (Tu-Tzeng-Chang) entropy $S^{\text{TTC}}_A$~\cite{Tu_2022}
      & $\mathcal{PT}$-transition 
      & $\mathcal{PT}$-unbroken case: oscillation.
          $\mathcal{PT}$-broken case: linear growth in free fermion systems and  exponential growth in interacting systems. \\
    \bottomrule
  \end{tabularx}
  \caption{Comparison of various quantities in the biothogonal quench dynamics of non-Hermitian systems. 
  When the system under a sudden quench by a Hamiltonian with parameters in the $\mathcal{PT}$-(un)broken phases, we
  refer the case to as the $\mathcal{PT}$-(unbroken) case. The $\mathcal{PT}$-transition means that the post-quench Hamiltonian undergoes the $\mathcal{PT}$-unbroken to $\mathcal{PT}$-broken transition by tuning parameters.
  The level-crossing indicates that the ground state of the post-quench Hamiltonian becomes the largest imaginary energy state deep in the $\mathcal{PT}$-broken phase,
  as heuristically presented in Figs.~\ref{Fig:1}(a-d))}
    \label{tab:entropy_comparison}
\end{table*}


This paper is organized as follows. In Sec.~\ref{Biorthogonal}, we introduce the framework of time evolution in the biorthogonal basis and analyze the quench dynamics of general structures of the biorthonormal density matrix in $\mathcal{PT}$-unbroken and $\mathcal{PT}$-broken cases. In Secs.~\ref{Yang_Lee} and \ref{Entanglement_Entropy}, we demonstrate the biorthogonal quench dynamics of observables and quantum geometry, and various entanglement quantities in the Yang-Lee model.
In Sec.~\ref{Free_interacting}, we discuss distinct quench dynamics of the TTC entropy in free-fermion systems and generic interacting systems. We conclude our results in Sec.~\ref{Conclusion}.


\section{Non-Hermitian biorthogonal quantum mechanics and its dynamics} 
\label{Biorthogonal}

For generic non-Hermitian quantum systems, the Hamiltonian can be expressed as 
\begin{equation}
    \hat{H} = \hat{H}_H + \hat{H}_{NH},
\end{equation}
where \(\hat{H}_H\) is the Hermitian component, and \(\hat{H}_{NH}\) introduces non-Hermitian effects. 
One can use the  biorthogonal basis to study the time evolution of non-Hermitian systems~\cite{Ashida2020,Brody2013}:  
The right (\(|R_n\rangle\)) and left (\(\langle L_n|\)) eigenstates of \(\hat{H}\) satisfy the eigenvalue equations~\cite{Brody2013}:
\begin{equation}
    \hat{H} |R_n\rangle = E_n |R_n\rangle, \quad \langle L_n| \hat{H} = \langle L_n| E_n^*.
\end{equation}  
Unlike in Hermitian systems, these eigenstates are generally non-orthogonal, 
one can consider using the biorthogonal formalism for a consistent time evolution framework~\cite{Brody2013,Ashida2020}.

The evolution of quantum states in the biorthogonal basis is governed by the propagators~\cite{Brody2013,wang2024}:  
\begin{equation}
    U_R(t) = e^{-i\hat{H}t}, \quad U_L(t) = e^{-i\hat{H}^\dagger t}.
\end{equation}  
Thus, the right and left wavefunctions evolve as:  
\begin{equation}
    |\psi_R(t)\rangle = U_R(t)|\psi_R(0)\rangle, \quad
    \langle \psi_L(t)| = \langle \psi_L(0)| U_L^\dagger(t).
\end{equation}  
Despite evolving independently, their inner product remains conserved~\cite{Ashida2020,Brody2013,wang2024}:  
\begin{equation}
    \langle \psi_L(t)|\psi_R(t)\rangle = \langle \psi_L(0)|\psi_R(0)\rangle = 1,
\end{equation}  
ensuring a well-defined probability framework. One can also rewrite the evolution of the state as the evolution of the (biorthogonal) density matrix $\rho^{RL} = |\psi_R \rangle \langle \psi_L |$.
In this expression, the dynamics of the density matrix satisfies the Heisenberg equation of motion, 
$\dot{\rho}^{RL}(t) = i [\rho^{RL}(t), H]$.
Similarly, the trace of the time-evolved biorthogonal density matrix \(\rho^{RL}(t)\) satisfies $ \text{Tr}(\rho_{RL}(t)) = 1$~\cite{Brody2013},
allowing the computation of physical observables and entanglement quantities using the biorthogonal density matrix in non-Hermitian systems.

The quench dynamics of these non-Hermitian quantum systems are strongly governed by eigenmodes of  \(\hat{H}\), where
long-time dynamics is dominated by eigenmodes with the largest imaginary components of \(E_n\), which amplify or suppress the wavefunction norm.
Previous study~\cite{wang2024} has shown this amplification/suppression can lead to an entanglement transition.
Their analysis is based on the (right-right) density matrix defined as 
\begin{align}
\rho^{RR} (t) =  \frac{|\psi_R(t) \rangle \langle \psi_R (t)|}{\langle \psi_R(t) | \psi_R(t) \rangle},
\label{Eq:rhoRR}
\end{align}
where the normalization of the density matrix is imposted~\cite{B_csi_2021,Turkeshi_2023}. 

In this paper, instead of using the (right-right) density matrix, we study the quench dynamics of physical observables and entanglement quantities in these
non-Hermitian using the biorthogonal density matrix. Specifically, we focus on the non-Hermitian systems with a $\mathcal{PT}$ symmetry.


\subsection{Quench dynamics following a sudden quench into the $\mathcal{PT}$-unbroken phase of the post-quench Hamiltonian: oscillations} \label{PT-Unbroken}
We consider the dynamics following a sudden quench into the $\mathcal{PT}$-unbroken phase of the post-quench Hamiltonian.
For a $\mathcal{PT}$-symmetric non-Hermitian Hamiltonian, in the $\mathcal{PT}$-unbroken phase, all the eigen-energies are real.
If the post-quench Hamiltonian is in the $\mathcal{PT}$-unbroken phase, some of the eigen-energies come in complex-conjugation pairs.
For the post-quench Hamiltonian in $\mathcal{PT}$-unbroken phase, the time evolution in the biorthogonal basis behaves similarly to that in Hermitian systems. No imaginary parts appear in the energy spectrum, meaning that eigenmodes undergo phase rotation rather than amplification or decay. Consequently, physical observables exhibit oscillatory behavior over time~\cite{Lu_2025, PhysRevLett.132.220402}, and no exponential divergence arises in either the state or its observables~\cite{Brody2013, Alice_2024,Lu_2025, PhysRevLett.132.220402}.

The time-dependent wavefunction in the right-basis takes the form
\begin{align}
    |\psi_R(t)\rangle 
    &= \sum_n c_n e^{-iE_n t}|R_n\rangle,
\end{align}
and similarly, in the left-basis
\begin{align}
    \langle\psi_L(t)|
    &= \sum_n c_n^* e^{iE_n t}\langle L_n|.
\end{align}

Since \( E_n \in \mathbb{R} \) in the $\mathcal{PT}$-unbroken case, the wavefunctions remain norm-preserving under the biorthogonal inner product, and no single eigenstate dominates the dynamics~\cite{Brody2013,li2022dynamicssimulationnumericalanalysis}. 
Since the left state and the right state are not dominated by any particular eigenstate, the density matrix remains oscillatory:
\begin{align}
    \rho^{RL}(t)
    &= |\psi_R(t)\rangle\langle \psi_L(t) |  \notag \\
    &= \sum\limits_{m,n}c_n c_m^* e^{-i(E_n - E_m)t}|R_n\rangle \langle L_m |.
\end{align}
The expectation value of a physical observable \( \hat{O} \) evolves as:
\begin{align}
    \langle \hat{O}(t) \rangle 
    &= \text{Tr}(\rho^{RL}\hat{O}) \notag \\
    &= \sum_{m,n} c_m^* c_n e^{i(E_m - E_n)t} \langle L_m | \hat{O} | R_n \rangle.
\end{align}

This behavior confirms that for a system suddenly quenched into the $\mathcal{PT}$-unbroken phase of the post-quench Hamiltonian, the biorthogonal time evolution preserves unitary-like features \cite{Brody2013, li2022dynamicssimulationnumericalanalysis, PhysRevB.109.224307}, despite the non-Hermitian nature of the Hamiltonian.
In addition, the reduced density matrix can be calculated by performing a partial trace of the density matrix, which remains oscillatory. As a result, the entanglement entropy oscillates over time. Moreover, the quantum geometry tensor is shown to oscillate over time since no single eigenstate dominates the dynamics. This point will be discussed in detail in ~\ref{QMT}.

 However, this picture changes drastically once the quench dynamics is governed by the $\mathcal{PT}$-broken case, where eigenvalues acquire nonzero imaginary parts and the dynamics become exponentially amplified. We now turn to an analysis of this exponential behavior in the $\mathcal{PT}$-broken case.

\subsection{Quench dynamics following a sudden quench into the \texorpdfstring{$\mathcal{PT}$}{PT}-broken phase of the post-quench Hamiltonian: exponential growth} \label{PT_Broken}
While the $\mathcal{PT}$-unbroken case is characterized by real energy spectra and bounded, oscillatory dynamics, a qualitative change occurs once the system is suddenly quenched into the $\mathcal{PT}$-broken phase of the post-quench Hamiltonian~\cite{wang2024, Alice_2024, li2022dynamicssimulationnumericalanalysis}. In this case, eigenvalues appear as complex-conjugate pairs, and the time evolution of the system is dominated by modes with nonzero imaginary parts. Although the initial state is composed of contributions from all eigenmodes, those with the largest imaginary components eventually dominate due to exponential amplification~\cite{wang2024, li2022dynamicssimulationnumericalanalysis,deng2024restoringkibblezurekscalingdefect}.

This exponential amplification fundamentally alters dynamics of the physical observables and the growth of the entanglement entropy. In particular, under biorthogonal time evolution, the mismatch between dominant right and left eigenstates gives rise to exponential terms in the density matrix  that do not affect probability conservation. These terms do not cancel in the $\mathcal{PT}$-broken case, leading to unbounded growth in observables and entanglement entropy. In the following sections, we characterize this exponential behavior in detail using both asymptotic analysis and numerical simulations.

\subsubsection{General structure}

The time evolution of the system can be analyzed using the eigenbasis expansion of the post-quench Hamiltonian. In the right-basis, the wavefunction evolves as:
\begin{align}
    |\psi_R(t)\rangle 
   \notag &= \sum_n c_n e^{-iE_n t}|R_n
    = \sum_n c_n e^{-i\text{Re}(E_n) t}e^{\text{Im}(E_n) t}|R_n\rangle \\
    &\sim c_{\max} e^{E_I t}|R_{\text{max}}\rangle.
\end{align}

Similarly, in the left-basis, the evolution follows:
\begin{align}
    \langle\psi_L(t)|
    \notag &= \sum_n c_n^* e^{iE_n t}\langle L_n| 
     = \sum_n c_n^* e^{i\text{Re}(E_n) t}e^{-\text{Im}(E_n) t}\langle L_n| \\
    &\sim c_{\min}^* e^{E_I t}\langle L_{\text{min}}|.
\end{align}
Here $|R_{\text{max}}\rangle$ and $\langle L_{\text{min}}| $ satisfy
\begin{align}
    \hat{H}|R_{\text{max}}\rangle &= (E_R + iE_I)|R_{\text{max}}\rangle, \\
   \langle L_{\text{min}}|  \hat{H} &= (E_R - iE_I)^*   \langle L_{\text{min}}| .
\end{align}
The biorthogonality leads to  $\langle L_{\text{min}}|R_{\text{max}}\rangle = 0$.
These results highlight a fundamental distinction between the orthogonal and biorthogonal formalisms. 
In the orthogonal basis, the eigenstate with the largest imaginary component in its eigenvalue eventually dominates the dynamics due to exponential amplification~\cite{wang2024,deng2024restoringkibblezurekscalingdefect,PhysRevB.109.224307}. 
In contrast, in the biorthogonal basis, the right and left components evolve asymmetrically: the dominant contributions arise from the eigenstates with the largest and smallest imaginary energies, respectively. 
This distinction between the  right-right density matrix $\rho^{RR}$ and the biorthogonal density matrix $\rho^{RL}$  leads to different dynamics~\cite{PhysRevB.109.224307}.

To make this structure more explicit, we analyze the left-right density matrix $\rho_{RL}$:
\begin{align}
    \rho^{RL}(t)
    &= |\psi_R(t)\rangle\langle \psi_L(t) |  \notag \\
    &= c_{\max} c_{\min}^* e^{2E_I t} \,  \big( |R_{\max} \rangle \langle L_{\min}| \big) \notag \\
    &\quad + e^{E_I t} \,  \bigg( c_{\max} \sum_{m\neq \min} c_m^* e^{iE_m t} |R_{\max} \rangle \langle L_m| \bigg) \notag \\
    &\quad + e^{E_I t} \,  \bigg( c_{\min}^* \sum_{n\neq \max} c_n e^{-iE_n t} |R_n \rangle \langle L_{\min}| \bigg) \notag \\
    &\quad + \sum_{\text{else}} c_n c_m^* e^{-i(E_n - E_m)t} \,  \big( |R_n\rangle \langle L_m| \big) \notag \\
    &\xrightarrow{t \to \infty}    P_0 + e^{E_I t} P_1 + e^{2E_I t} P_2.
    \label{Eq:rhoRL}
\end{align}
Here we have
\begin{align}
    \text{Tr}(P_0) = 1, \, \text{Tr}(P_1) = 0, \, \text{Tr}(P_2) = 0,
\end{align}
because of the trace preserving property $ \rm Tr \rho^{RL}(t)=1$. 






On the other hand,  for the right-right basis definition of the density matrix $\rho^{RR}(t)$, one works solely with right eigenstates and their Hermitian conjugates.  
To preserve normalization at each time step, the state must be renormalized continuously [see Eq.~(\ref{Eq:rhoRR})].
For the $\mathcal{PT}$-broken case, the post-quench wavefunction is dominated by the right eigenmode with largest imaginary eigenenergy and 
has an exponential growth form $|\psi_R(t)\rangle=|\psi_R(t)\rangle \sim  e^{E_I t}$. 
However, the denominator in Eq.~(\ref{Eq:rhoRR}) also has an exponential growth form which cancels the exponential growth form of the numerator.
The long-time behavior of the $\rho^{RR}(t)$ has the form 
\begin{align}
\rho^{RR}(t) \xrightarrow{t \to \infty} |R_{{\max}}\rangle\langle R_{{\max}}|.
\end{align}



Comparing Eq.~(\ref{Eq:rhoRL}) and Eq.~(\ref{Eq:rhoRR}), $\rho^{RL}(t\to\infty)$ retains off-diagonal structure with an overall \(e^{2E_I t}\) scaling and $\rho^{RL}(t\to\infty)$ has no dynamics.
This distinct feature gives rise to different post-quench dynamics in entanglement entropy.
We discuss these results in detail in Sections~\ref{QMT} and~\ref{Entanglement_Entropy}.  

\section{Biorthogonal dynamics in the Yang-Lee model}
\label{Yang_Lee}

To illustrate the contrasting dynamical behaviors of the $\mathcal{PT}$-unbroken and $\mathcal{PT}$-broken cases, we take the one-dimensional Yang-Lee model as a concrete example.

\subsection{The Model}
\begin{figure}[htbp]
    \centering
    \begin{subfigure}[t]{0.39\linewidth}
        \centering
        \adjustbox{valign=t}{\includegraphics[width=\linewidth]{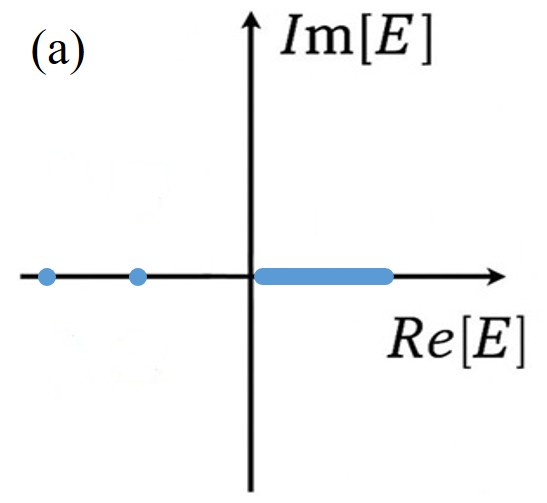}}
    \end{subfigure}
    \begin{subfigure}[t]{0.39\linewidth}
        \centering
        \adjustbox{valign=t}{\includegraphics[width=\linewidth]{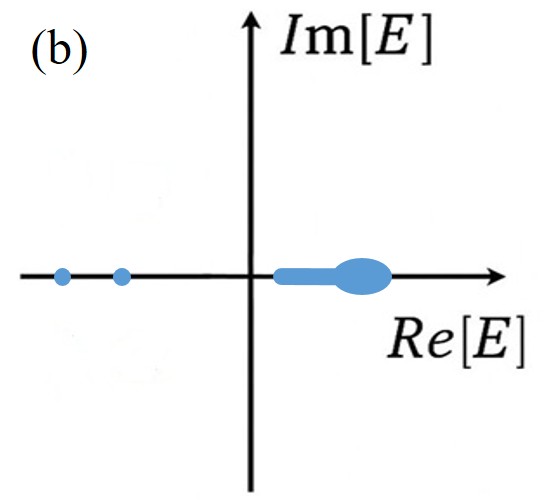}}
    \end{subfigure}
       \begin{subfigure}[t]{0.39\linewidth}
        \centering
        \adjustbox{valign=t}{\includegraphics[width=\linewidth]{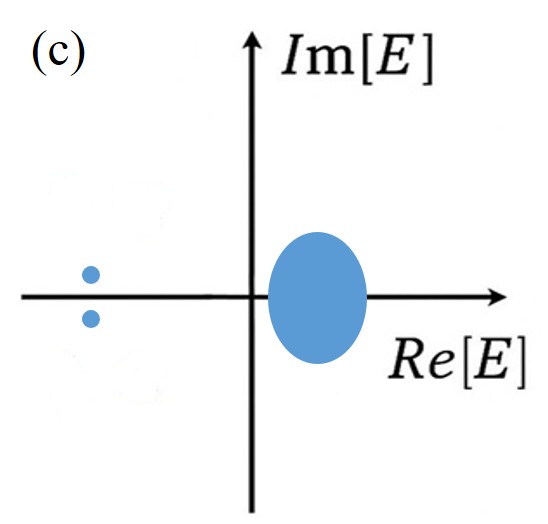}}
    \end{subfigure}
    \begin{subfigure}[t]{0.39\linewidth}
        \centering
        \adjustbox{valign=t}{\includegraphics[width=\linewidth]{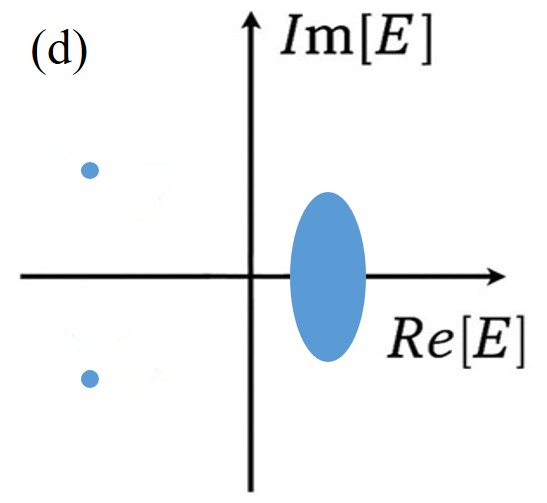}}
    \end{subfigure}
    \hfill
    \begin{subfigure}[t]{0.8\linewidth}
        \centering
        \adjustbox{valign=t}{\includegraphics[width=\linewidth]{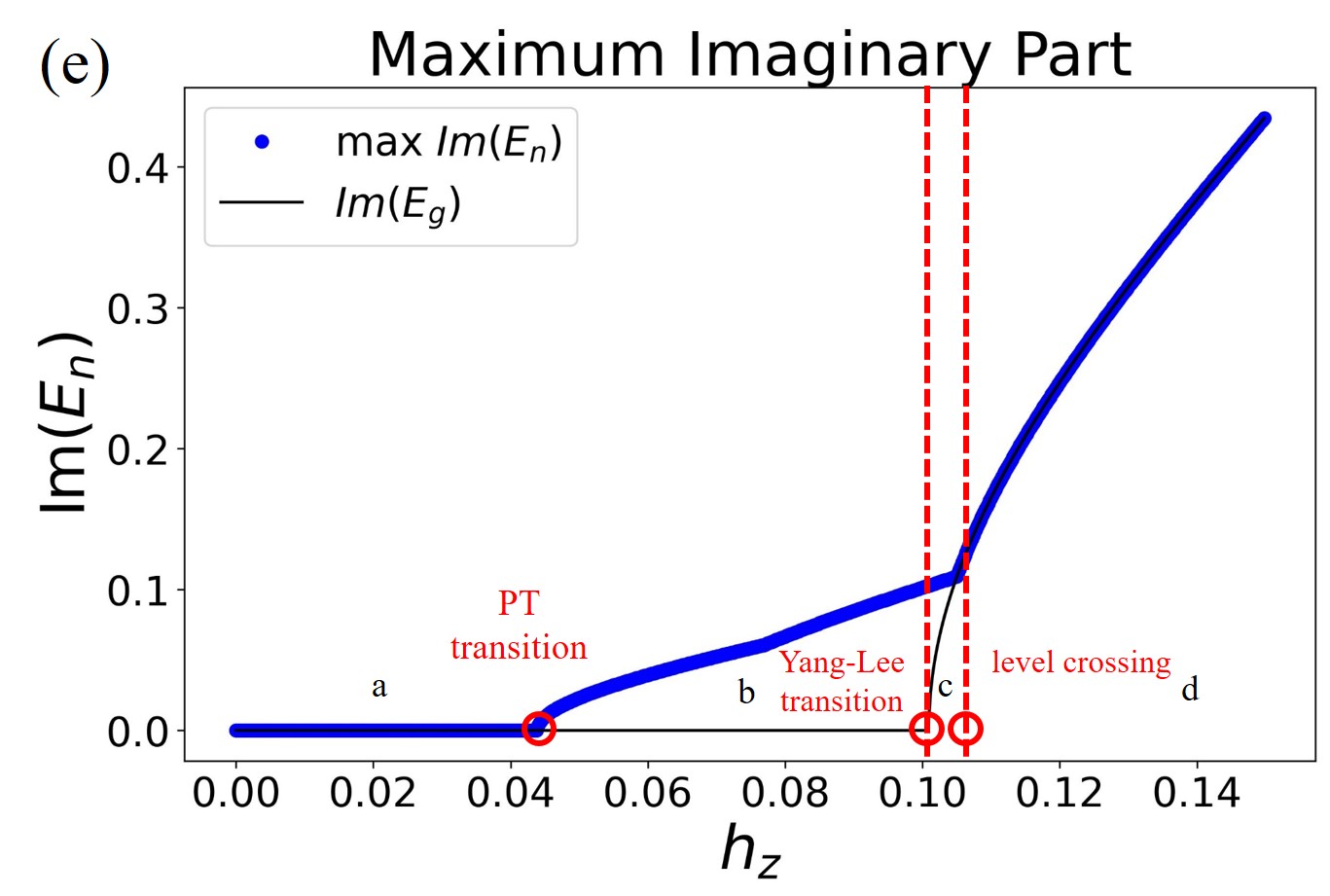}}
    \end{subfigure}
  \caption{Schematic evolution of the Yang-Lee model spectrum in the complex-energy plane as $|h_z|$ increases. (a) Before the $\mathcal{PT}$-transition, all eigenvalues lie on the real axis. (b) For $h_{PT}<|h_z|<h_{YL}$, higher-energy modes acquire nonzero imaginary parts while the ground and first-excited pair remains real.  (c) $h_{YL}<|h_z|<h_{c}$, the ground and first-excited levels coalesce at the Yang-Lee transition and split into complex conjugate pairs, but do not yet possess the largest $|\mathrm{Im}\,E|$. (d) Beyond the level-crossing field $h_c$, the ground and first-excited pair becomes the dominant conjugate pair with the maximal imaginary component, driving exponential amplification. (e) Maximum imaginary part of the energy spectrum in the Yang-Lee model. Here we set $(J_1, J_2,h_x)=(0.4, 0.1, 0.9)$ in Eq.~(\ref{eq:YL_Hamiltonian}).
  }  
  \label{Fig:1}
\end{figure}

To illustrate our general framework, we study the $(1+1)$-dimensional Yang-Lee model~\cite{Yang1952, Lee1952, Fisher1978, Cardy1985, Lu_2025, Jian2021,Zhai_2020, yan2024dissipativedynamicalphasetransition}, defined by the spin-\(1/2\) chain with both real transverse and imaginary longitudinal fields:
\begin{align}
  H_{\mathrm{YL}}
  =&-J_1 \sum_{j=1}^L \sigma_j^z \sigma_{j+1}^z
    -J_2 \sum_{j=1}^L \sigma_j^z \sigma_{j+2}^z
    - h_x \sum_{j=1}^L \sigma_j^x   \notag \\    
    &- i\,h_z \sum_{j=1}^L \sigma_j^z,
  \label{eq:YL_Hamiltonian}
\end{align}
where \(\sigma_j^{x,z}\) are Pauli matrices on site \(j\), \(J_1>0\) is the Ising coupling, \(h_x\) is the real transverse field, and \(h_z\) is the imaginary longitudinal field.  
Here we include the second nearest-neighbor coupling $J_2$ for numerical efficiency to allow faster simulations. In the absence of external field $h_x=h_z=0$,
the term $J_2$ breaks the integrability. We will focus on the case where $J_1 > J_2 > 0$. 
Under the combined action of spatial inversion \(P: \sigma_x\to -\sigma_x\) and time reversal \(T: i\to -i\), \(H_{\mathrm{YL}}\) is $\mathcal{PT}$-symmetric.  For \(|h_z|<h_{\mathcal{PT}}\), all eigenvalues are real ($\mathcal{PT}$-unbroken phase), whereas at \(|h_z|=h_{\mathcal{PT}}\) two levels coalesce (for high energy modes) and for \(|h_z|>h_{\mathcal{PT}}\) they split into complex-conjugate pairs [See Figs.~\ref{Fig:1}(a) and (b)]. 
This transition is referred to as the $\mathcal{PT}$-transition.
When the ground state and the first excited state merge into the same energy level and split into a complex-conjugate pair, the transition is referred to as the Yang-Lee transition [Fig.~\ref{Fig:1}(c)].
When the splitting of the ground state and the first excited state along the imaginary axis exceeds the largest and the smallest imaginary part of the other energies of the other eigenmodes,
the level-crossing occurs  [Fig.~\ref{Fig:1}(d)]~\cite{PhysRevB.104.155141}.
Different transitions can be visualized by tracking the largest imaginary energy as shown in Fig.~\ref{Fig:1}(e).
Now let us discuss the biorthogonal quench dynamics of this model in terms of the observable quantities~\cite{e27020170,yu2024emergentdynamicalquantumphase}, the quantum geometry, and the different definitions of entanglement entropies~\cite{soares2025symmetriesconservationlawsentanglement}.

\subsection{Physical Observable}

For a physical observable \(\hat{O}\), its expectation value evolves as:
\begin{align}
    \langle \hat{O} (t)\rangle \notag
    &={ \rm Tr} (    \rho_{RL}(t) \hat{O}) \\ 
    &= \sum\limits_{n,m} c_n c_m^* e^{-i(E_n - E_m)t} \langle L_m | \hat{O} | R_n \rangle.
\end{align}

If \(\hat{O}\) commutes with \(\hat{H}\), the equation of motion $d\hat{O}(t)/dt = i   [\hat{H},\hat{O}] = 0$,
indicating   $\langle \hat{O} (t)\rangle$ is a constant over time.





\begin{figure}[htbp]
  \begin{subfigure}[b]{0.49\linewidth}
    \includegraphics[width=\linewidth]{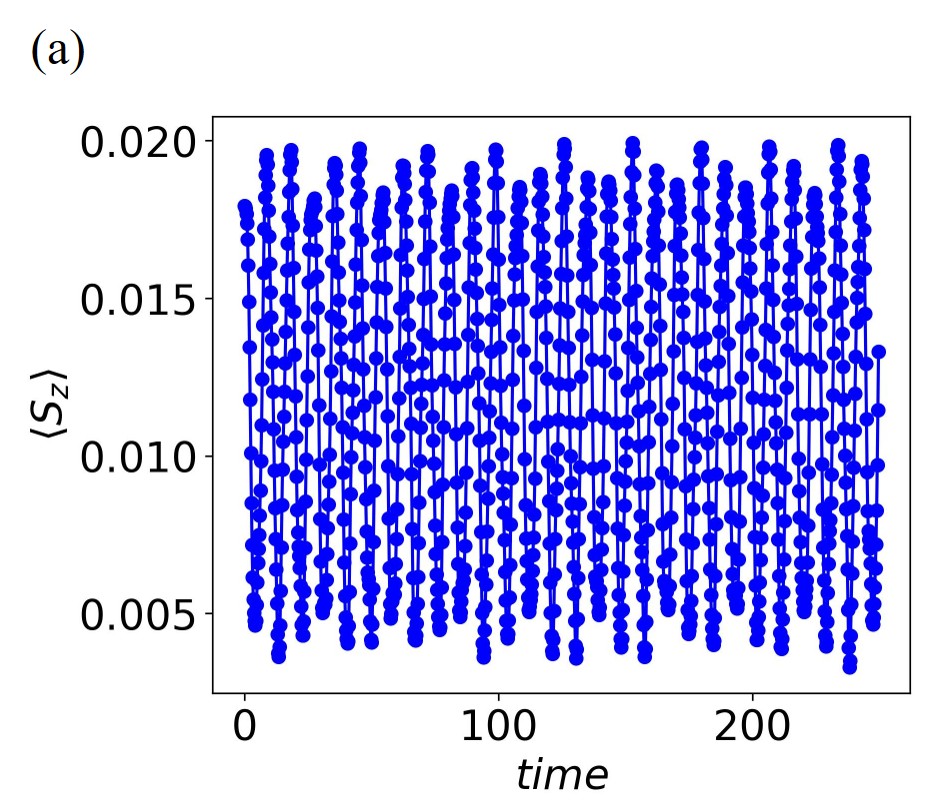}
  \end{subfigure}%
  \hfill
  \begin{subfigure}[b]{0.49\linewidth}
    \includegraphics[width=\linewidth]{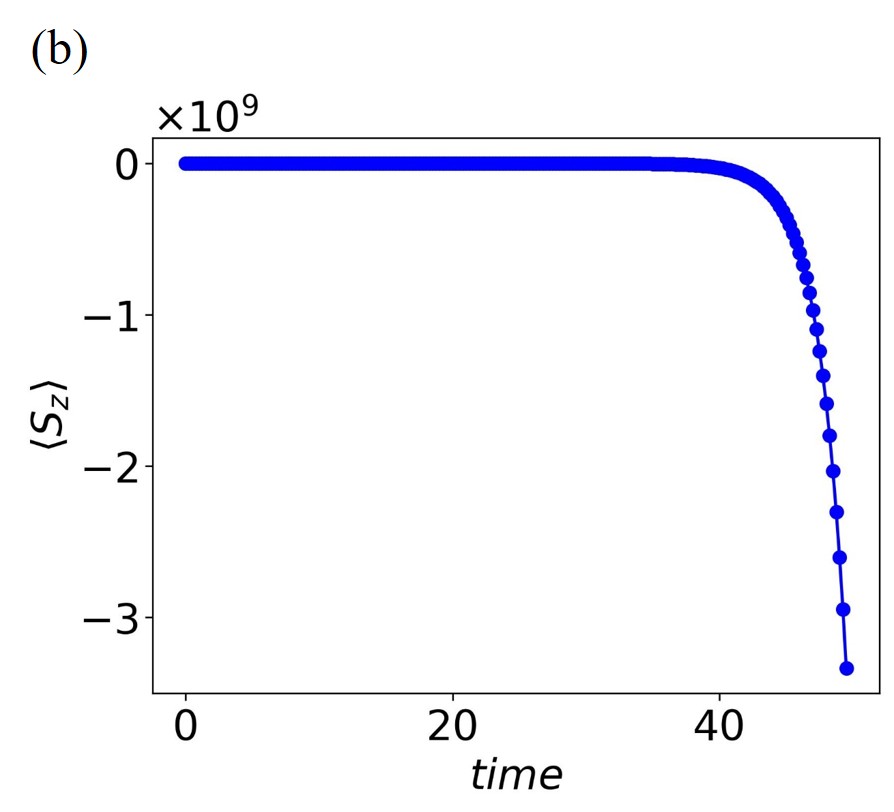}
  \end{subfigure}
  \caption{(a) The $\mathcal{PT}$-unbroken case: the observable \(S_z(t)\) oscillates in time where the parameters are $(J_1, J_2,h_x, h_z)=(0.4, 0.1, 0.9,0.03)$. (b) The $\mathcal{PT}$-broken case: \(S_z(t)\) (negative) grows exponentially with rate \(e^{E_I t}\), demonstrating the exponential amplification in $\mathcal{PT}$-broken case where the parameters are $(J_1, J_2,h_x, h_z)=(0.4, 0.1, 0.9,0.12)$.}
  \label{Fig:2}
\end{figure}

For $ [\hat{H},\hat{O}] \neq 0$, the expectation value of $\hat{O}$
is oscillatory if the post-quench Hamiltonian is $\mathcal{PT}$-unbroken and (negative) grows exponentially  if the post-quench Hamiltonian is $\mathcal{PT}$-broken.
We compute the $  \langle \hat{S_z} \rangle$ for the Yang-Lee model as shown in  Fig.~\ref{Fig:2}. 


\subsection{Quantum Geometry in Non-Hermitian Systems}
\label{QMT}
Quantum geometry has emerged as a powerful framework for characterizing the geometric and topological features of quantum states, and has recently been extended to non‐Hermitian systems, where it can diagnose topological phase transitions via singular behavior in the metric tensor~\cite{Ye2024,behrends2025quantumgeometrynonhermitiansystems}. Unlike Hermitian systems, where the quantum metric tensor (QMT) is always real and symmetric, non-Hermitian systems require a more generalized approach due to the biorthogonal nature of their eigenstates. The non-Hermitian quantum metric tensor (NH-QMT) is defined as~\cite{Ye2024, behrends2025quantumgeometrynonhermitiansystems}:

\begin{equation}
\chi_{\mu\nu}^{RL} = \langle \partial_\mu \psi_R | \partial_\nu \psi_L \rangle 
- \langle \partial_\mu \psi_R | \psi_L \rangle \langle \psi_R | \partial_\nu \psi_L \rangle,
\end{equation}

where \(|\psi_R\rangle\) and \(|\psi_L\rangle\) represent the right and left eigenstates of the non-Hermitian Hamiltonian, and \(\partial_\mu\) and \(\partial_\nu\) denote derivatives with respect to parameters \(\mu\) and \(\nu\) in the system. 

To explore whether quantum geometry can similarly detect dynamical phase transitions in quench protocols, we apply the NH–QMT to the Yang–Lee model. In this model, the NH-QMT is defined for the parameter space spanned by time (\(t\)) and the imaginary external field (\(h_z\)).  The four elements of the NH-QMT in this parameter space are:

\begin{align}
\chi_{h_z h_z}^{RL} &= \langle \partial_{h_z} \psi_L | \partial_{h_z} \psi_R \rangle 
- \langle \partial_{h_z} \psi_L | \psi_R \rangle \langle \psi_L | \partial_{h_z} \psi_R \rangle,  \notag \\
\chi_{h_z t}^{RL} &= \langle \partial_{h_z} \psi_L | \partial_t \psi_R \rangle 
- \langle \partial_{h_z} \psi_L | \psi_R \rangle \langle \psi_L | \partial_t \psi_R \rangle,  \notag  \\
\chi_{t h_z}^{RL} &= \langle \partial_t \psi_L | \partial_{h_z} \psi_R \rangle 
- \langle \partial_t \psi_L | \psi_R \rangle \langle \psi_L | \partial_{h_z} \psi_R \rangle,    \notag  \\
\chi_{t t}^{RL} &= \langle \partial_t \psi_L | \partial_t \psi_R \rangle 
- \langle \partial_t \psi_L | \psi_R \rangle \langle \psi_L | \partial_t \psi_R \rangle.
\end{align}

The NH-QMT provides a geometric measure of the parameter space's curvature, offering insights into how quantum states evolve under changes in \(t\) or \(h_z\). Since we perform a quench dynamic, the pre-quench state is independent of the quench Hamiltonian. The quantum geometry has the following analytical solutions:

\begin{align}
    |\partial_t \psi_R (t) \rangle &= 
    (\partial_t e^{-i\hat{H}t})|\psi_R (0) \rangle
    = -i\hat{H} | \psi_R (t) \rangle, \notag \\
    |\partial_{h_z} \psi_R (t) \rangle &= 
    (\partial_{h_z} e^{-i\hat{H}t})|\psi_R (0) \rangle
    = -i\frac{\partial \hat{H}}{\partial h_z}t  | \psi_R (t) \rangle.
\end{align}

\begin{figure}[htbp]
\begin{subfigure}[b]{0.48\linewidth}        
    \centering
    \includegraphics[width=\linewidth]{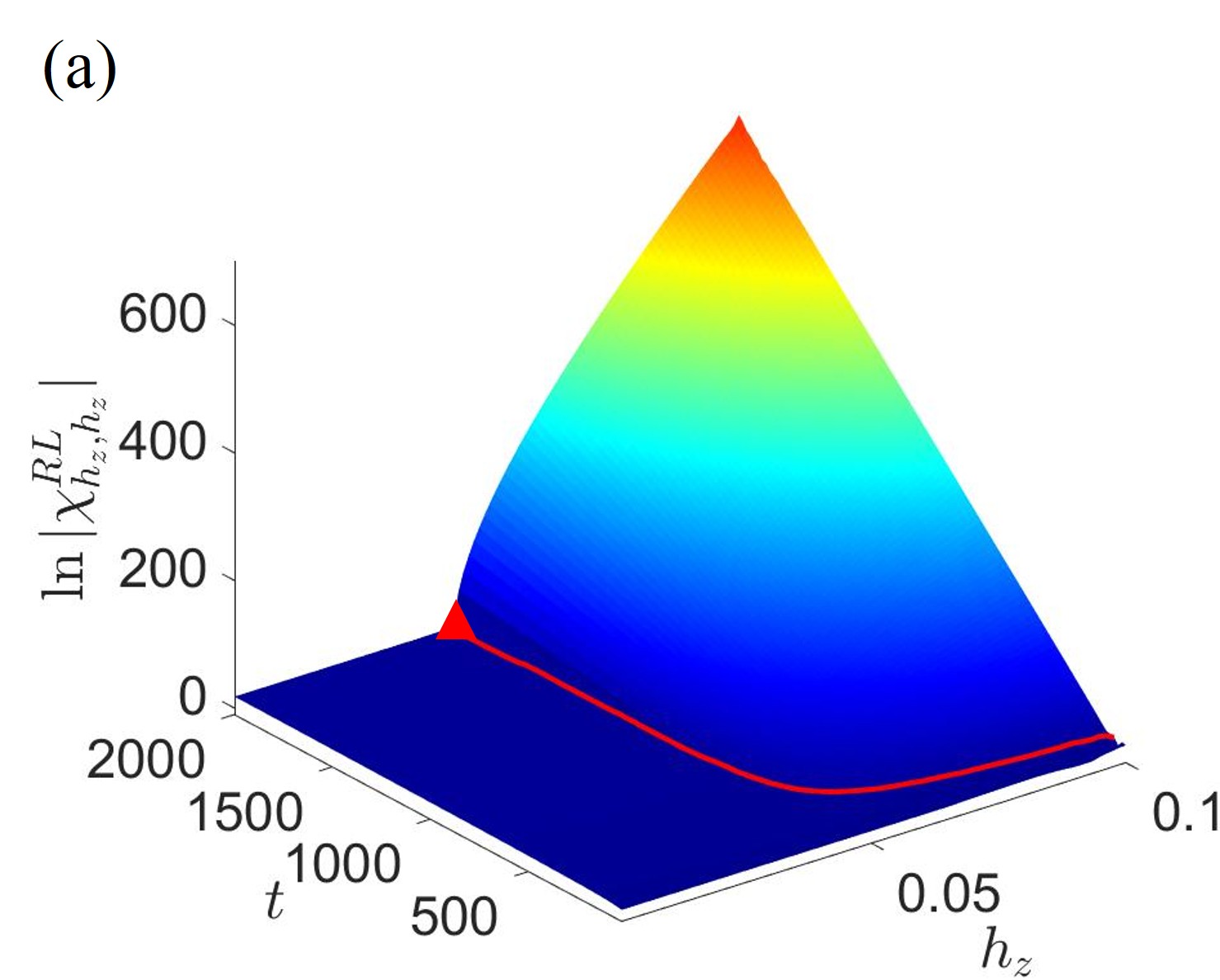}
\end{subfigure}
\begin{subfigure}[b]{0.48\linewidth}     
    \centering
    \includegraphics[width=\linewidth]{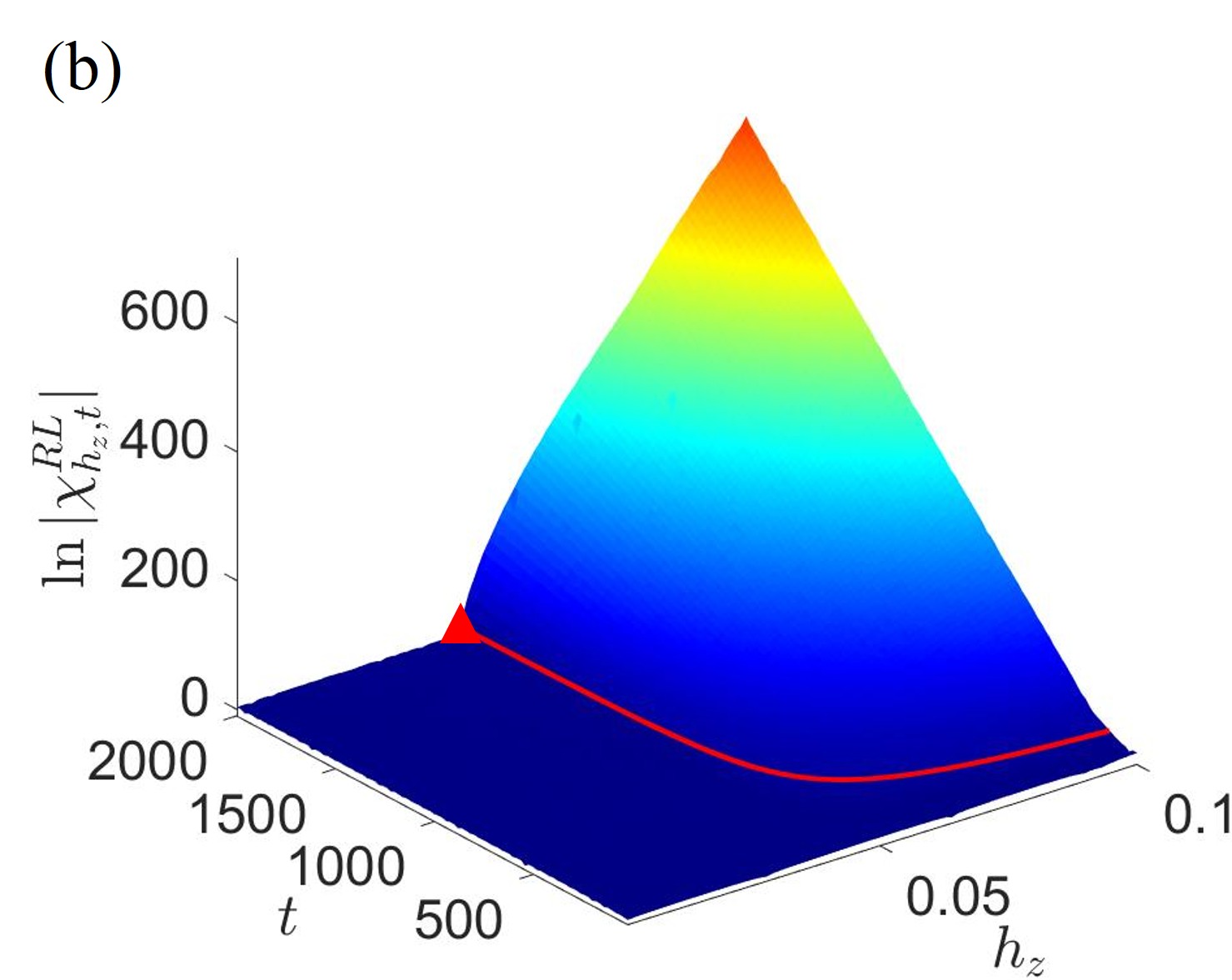}
\end{subfigure}
\begin{subfigure}[b]{0.48\linewidth}   
    \centering
    \includegraphics[width=\linewidth]{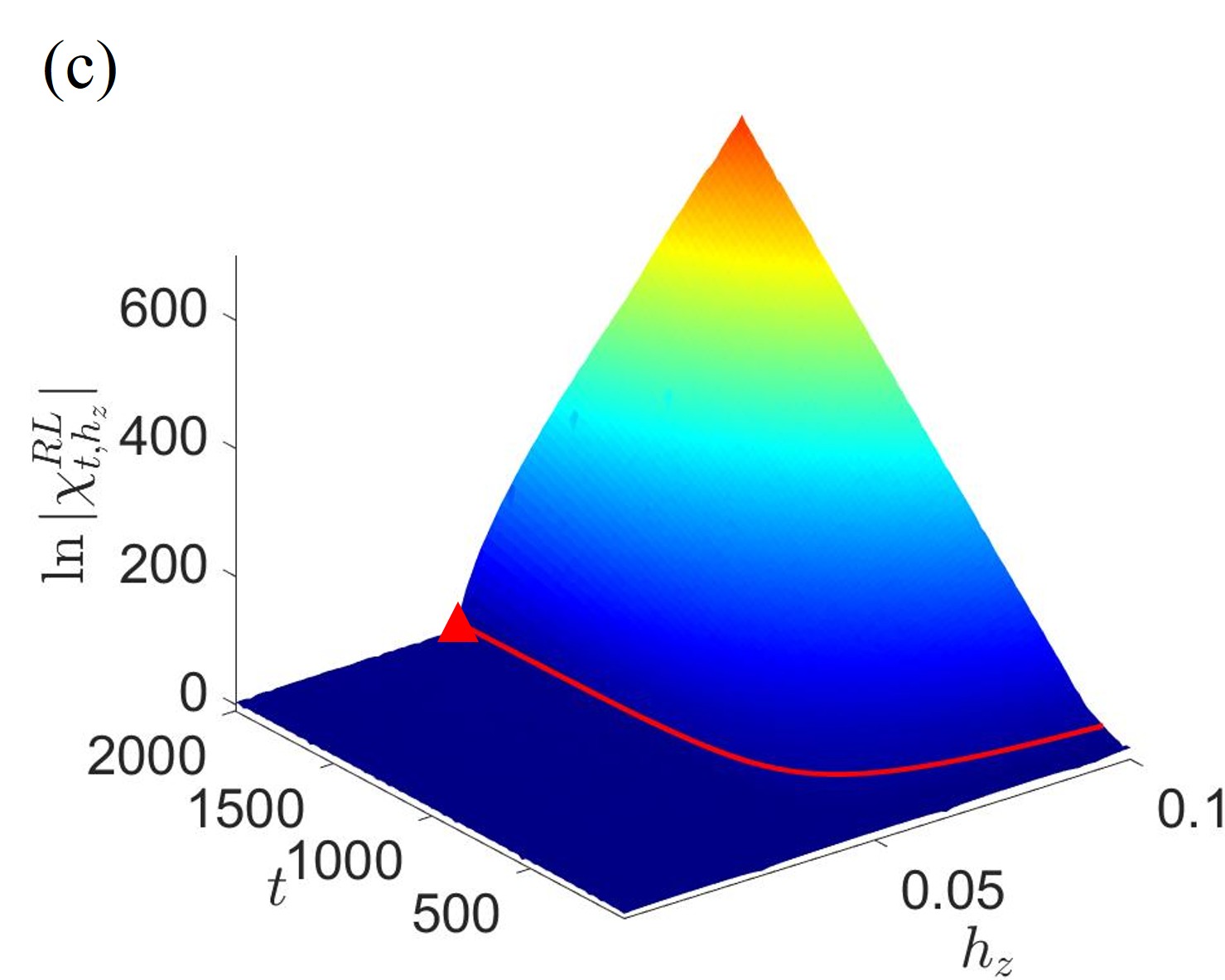}
\end{subfigure}
\begin{subfigure}[b]{0.48\linewidth}       
    \centering
    \includegraphics[width=\linewidth]{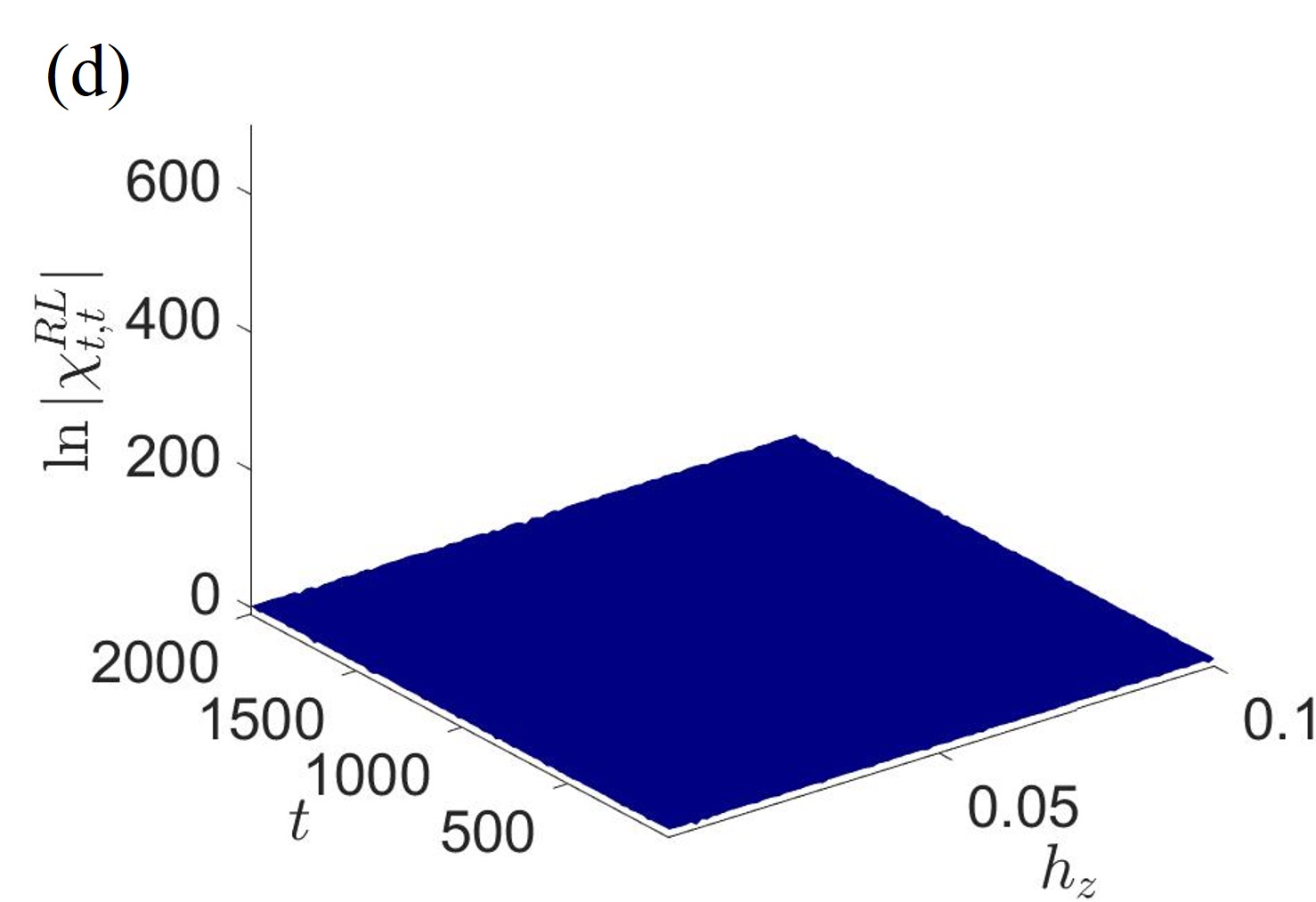}
\end{subfigure}
\caption{The elements of the quantum metric capture the $\mathcal{PT}$-transition points of the energy spectrum, denoted by the red triangles in the quantum geometric plot.
The red line indicates the critical time $t_c =  {\rm const.} /  ( h_z - h_{\mathcal{PT}})$. The elements $\chi_{h_z h_z}^{RL}$, $\chi_{h_z t}^{RL}$, and $\chi_{t h_z}^{RL}$ exhibit exponential growth along the $t$ direction in the $\mathcal{PT}$-broken phase, whereas $\chi_{t t}^{RL}$ does not grow exponentially.}
\label{fig:quantum_geometry}
\end{figure}

For the post-quench Hamiltonian being $\mathcal{PT}$-unbroken, the four elements of quantum metrics remain constant over time, exhibiting linear and quadratic growth. 
However, once the post-quench Hamiltonian transits to a $\mathcal{PT}$-broken phase, the off-diagonal terms of the quantum metrics begin to grow exponentially. 
This result is due to the non-commutative property 
$[\hat{H},\frac{\partial\hat{H}}{\partial h_z}] \neq 0$,
which gives rise to an exponentially growing term for the $\mathcal{PT}$-broken case. 
For the Yang-Lee model, as shown in Fig.~\ref{fig:quantum_geometry}, the elements $ \chi_{h_z h_z}^{RL}, \chi_{h_z t}^{RL},$ and  $\chi_{t h_z}^{RL} $ exhibit 
exponential growth along the \( t \) direction, when the post-quench Hamiltonian is in the $\mathcal{PT}$-broken phase. 
Since the maximum imaginary part of the eigenenergy $E_I \propto (h_z - h_{\mathcal{PT}})$, we can define a characteristic time $t_c =  {\rm const.} /  ( h_z - h_{\mathcal{PT}})$
which separates distinct features of $ \chi_{h_z h_z}^{\alpha, \beta}$ as shown in the red line in Figs.~\ref{fig:quantum_geometry}(a-c).
On the other hand, due to $[H,H]=0$, $ \chi_{t t}^{RL} $ does not evolve over time [see Fig.~\ref{fig:quantum_geometry}(d)].
The dynamical property of the NH-QMT is similar to the dynamical property of observable quantities, indicating that the quench dynamics of the NH-QMT can detect the $\mathcal{PT}$-transition.

\section{Different entanglement entropies}
\label{Entanglement_Entropy}

In non-Hermitian systems, the biorthogonal density matrix $\rho^{RL} = |\psi_R \rangle \langle \psi_L |$ is non-Hermitian.
The non-Hermiticity leads to non-positivity of the biothogonal reduced density matrix (RDM) $\rho_A^{RL} = {\rm Tr}_B \rho^{RL}$, which causes entanglement entropy to be complex and depends on the choice of the branch cuts.
In Ref.~\cite{Tu_2022}, a \textit{generalized entanglement entropy} was introduced to resolve this problem. Here we referred to it as the Tu-Tzeng-Chang (TTC) entropy~\cite{Tu_2022, Fossati_2023, Yang_2024}, which is defined as:
\begin{equation}
    S^{\text{TTC}}_A = -\text{Tr}(\rho_A^{RL} \ln |\rho_A^{RL}|).
\end{equation}
Here $|\rho_A^{RL}|$ means we take the absolute value of the eigenvalues of $\rho_A^{RL}$.
This definition ensures that the entanglement entropy remains real 
throughout the evolution, even when the RDM contains complex eigenvalues.
The TTC entropy has been shown to capture critical behaviors such as logarithmic scaling and negative central charges in non-unitary conformal field theories~\cite{Chang_2020, Tu_2022, Hsieh2023}.
The corresponding spectrum of the RDM was also investigated recently, which is considered as the entanglement Hamiltonian of a non-Hermitian systems~\cite{Rottoli2024, Yang_2024}.

On the other hand, one can also define the singular-value-decomposition (SVD) entanglement entropy~\cite{Loc2019, Herviou_2019, Parzygnat_2023}, where the SVD values of RDM are always positive ensuring that
the SVD entanglement entropy is real.
First, we need to construct a normalized SVD-RDM from the RDM $\rho_A^{RL}$~\cite{Parzygnat_2023}:
\begin{equation}
  \rho_A^{1|2}
  \;=\;
  \frac{\sqrt { (\rho_A^{RL})^\dagger \rho_A^{RL}}}
       {\text{Tr}\!\Bigl[\sqrt{(\rho_A^{RL})^\dagger \rho^{RL}_A} \Bigr]}.
  \label{eq:svd_rho}
\end{equation}
Then we can define its entanglement entropy~\cite{Parzygnat_2023}:
\begin{equation}
  S\bigl(\rho_A^{1|2}\bigr)
  = -\,\text{Tr} \! \bigl[\rho_A^{1|2}\,\ln\rho_A^{1|2}\bigr].
  \label{eq:svd_entropy}
\end{equation}

Last, as studied previously, the right-right density matrix $\rho_{RR}(t)$ can be considered as a quantum state subject to continuous measurements~\cite{Jian_2021, Jian2021}. At any instant time $t$,
the quantum state needs to be normalized as given in Eq.~(\ref{Eq:rhoRR}). The corresponding entanglement entropy has the usual definition $S_A= -{\rm Tr} \rho^{RR}_A \ln \rho_A^{RR}$,
where $\rho_A^{RR} ={\rm Tr}_B \rho^{RR}$. The measurement-induced entanglement transition describes  how quantum systems transition between different entanglement phases due to the influence of local measurements. 
In Ref.~\cite{Jian2021}, the entanglement entropy (right-right basis) of the long-time steady states have distinct features under the Yang-Lee like Hamiltonian, where the transition is triggered by a level crossing.

The TTC entropy $S^{\rm TTC}_A$, the SVD entanglement entropy $S\bigl(\rho_A^{1|2}\bigr)$, and the entanglement entropy $S_A$, have different quench dynamics as summarized in Table~\ref{tab:entropy_comparison}.
For the $\mathcal{PT}$-unbroken case, the quench dynamics of all the entropies have oscillatory behaviors as shown in Figs.~\ref{fig:Yang_Lee_entropy} (a-c).
On the other hand, for the post-quench Hamiltonian in the $\mathcal{PT}$-broken phase, the quench dynamics of different the entropies have distinct behaviors as discussed in the follows.

\subsection{Long-time dynamics of TTC entropy in the \texorpdfstring{$\mathcal{PT}$}{PT}-broken case}

Now, we consider the TTC entropy computed from a biorthogonal reduced density matrix evolved in time. The biorthogonal reduced density matrix at long time can be expressed as
\begin{align}
    \rho^{RL}_A(t)
    \xrightarrow{t \to \infty} P^{A}_0 + e^{E_I t} P^{A}_1 + e^{2E_I t} P^{A}_2,
\end{align}
where $P^{A}_i = {\rm Tr}_B P_i$ with $i=0,1,2$.
The trace preserving of the $ \rho^{RL}_A(t)$ leads to ${\rm Tr}_A P^{A}_0 = 1$ and ${\rm Tr}_A P^{A}_1 = {\rm Tr}_A P^{A}_2 = 0$.

The TTC entropy at long time is
\begin{align}
S^{\rm TTC}_A =& -\text{Tr} \big(\rho^{RL}_A \ln |\rho_A^{RL}| \big) \notag \\
= &-2E_I t  - \text{Tr}\bigg( \big[ P^{A}_0 + e^{E_1 t} P^{A}_1 + e^{2E_1 t} P^{A}_2 \big]  \notag\\
& \times \ln \big| e^{-2E_I t}(P^{A}_0 \quad + e^{E_I t} P^{A}_1 + e^{2E_I t} P^{A}_2) \big| \bigg).
\end{align}

In long-time limit, $e^{2 E_It} P^A_2$ is the dominate term  and the TTC entropy has the form 

\begin{align}
S^{TTC}_A \xrightarrow{t \to \infty} -2 E_It - e^{2E_1 t} \text{Tr}( P^{A}_2 \ln | P^{A}_2 | ) .
\label{Eq:STT_int}
\end{align}

This result demonstrates that for the $\mathcal{PT}$-broken case, the TTC entropy grows exponentially [see Figs.~\ref{fig:Yang_Lee_entropy}(f)].
On the other hand, for the entanglement entropy $S_A(t)$ constructed from the $\rho^RR(t)$, the long-time dynamics of $S_A(t)$ is a constant  [see Figs.~\ref{fig:Yang_Lee_entropy}(d)].



\subsection{Long-time dynamics of SVD entropy in the \texorpdfstring{$\mathcal{PT}$}{PT}-broken case}
\begin{figure*}[t]  
\begin{subfigure}[b]{0.3\linewidth}        
     \includegraphics[width=\linewidth]{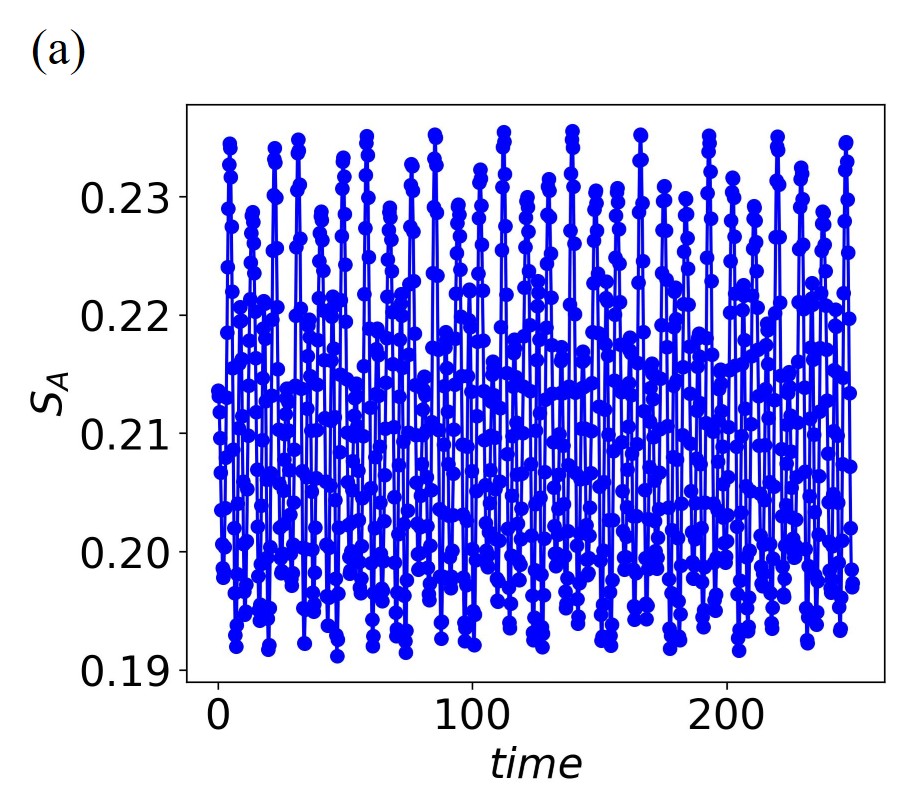}
\end{subfigure}
\begin{subfigure}[b]{0.3\linewidth}        
     \includegraphics[width=\linewidth]{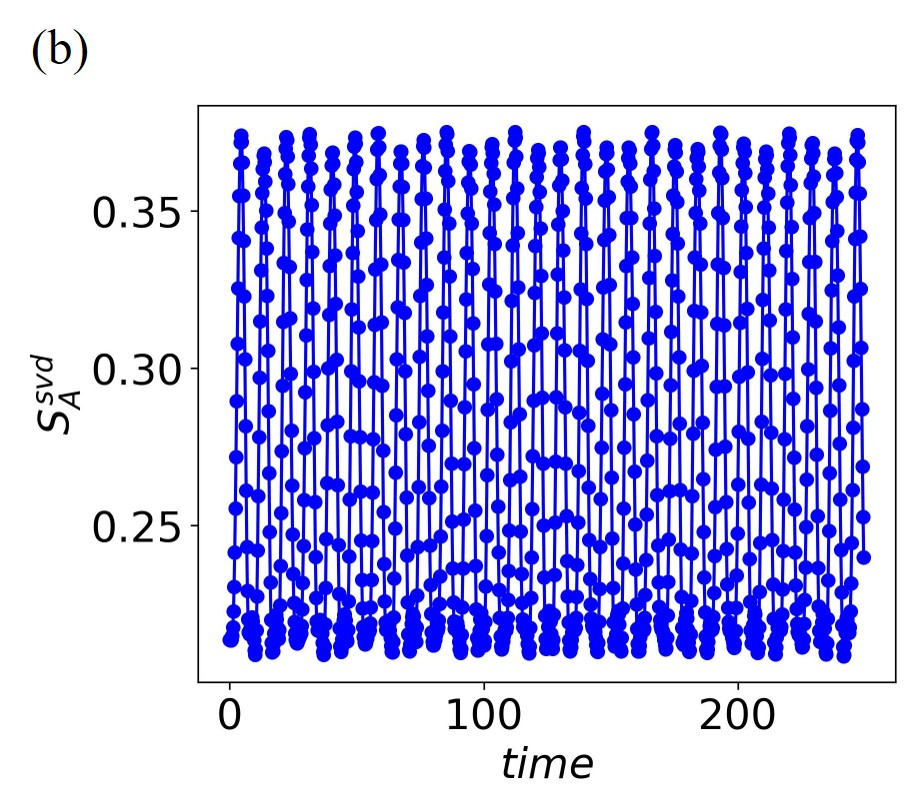}
\end{subfigure}
\begin{subfigure}[b]{0.3\linewidth}        
     \includegraphics[width=\linewidth]{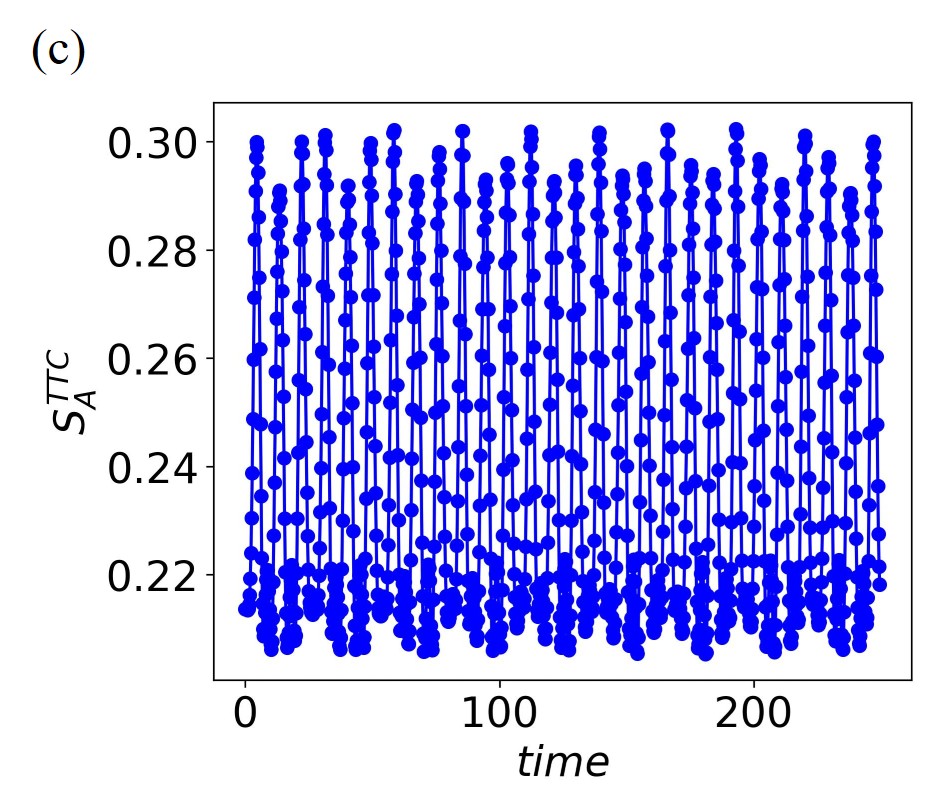}
\end{subfigure}
\begin{subfigure}[b]{0.3\linewidth}        
     \includegraphics[width=\linewidth]{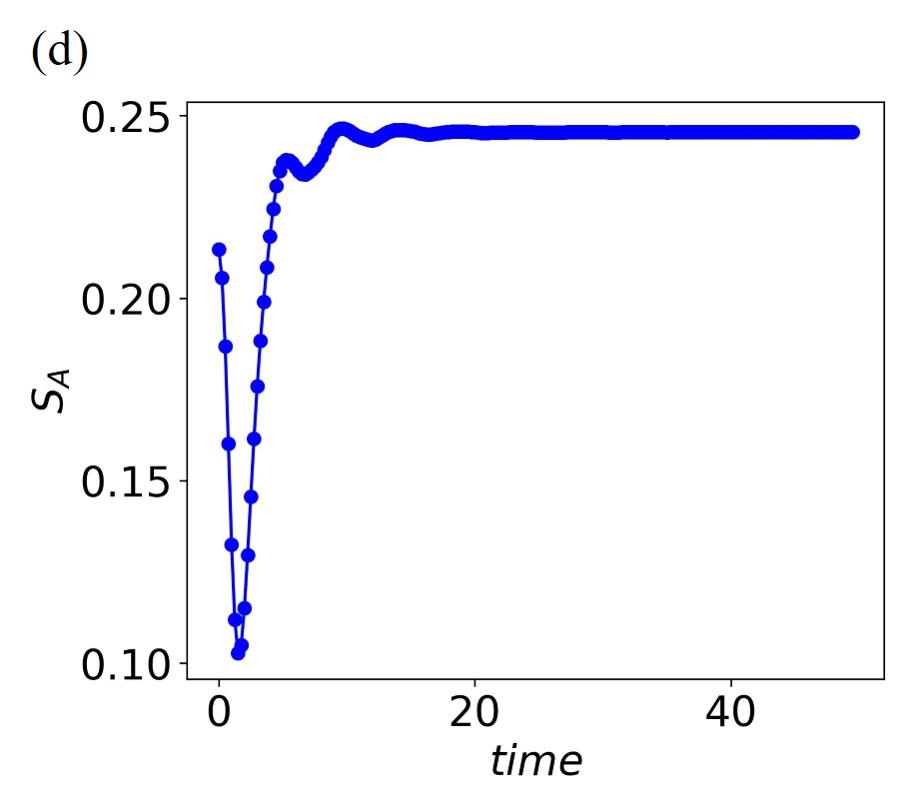}
\end{subfigure}
\begin{subfigure}[b]{0.3\linewidth}        
     \includegraphics[width=\linewidth]{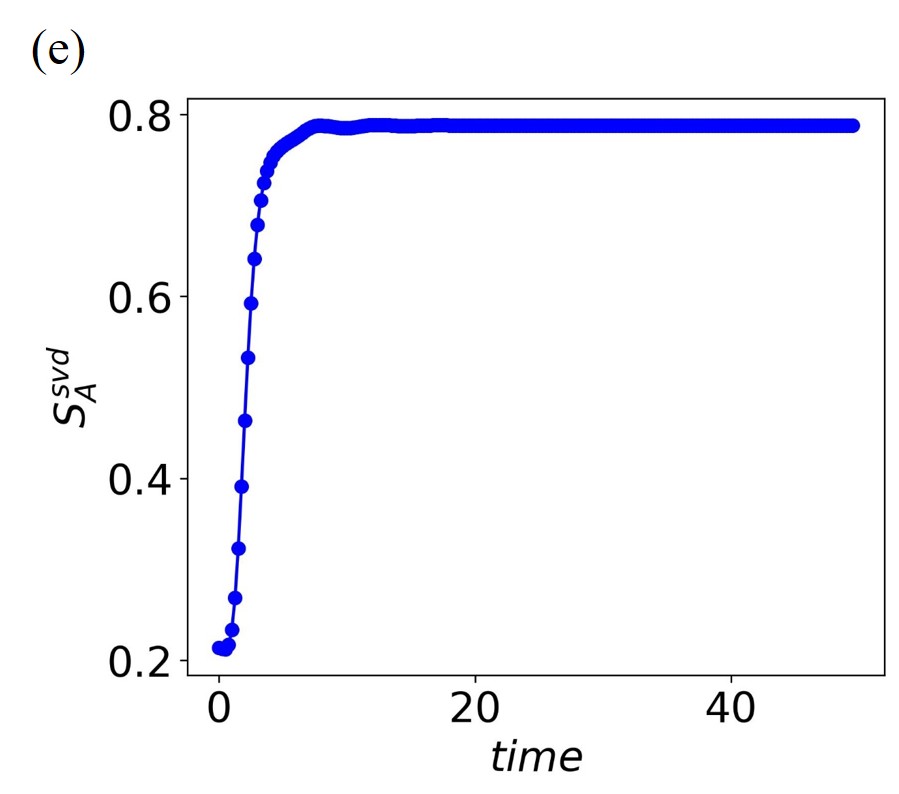}
\end{subfigure}
\begin{subfigure}[b]{0.3\linewidth}        
     \includegraphics[width=\linewidth]{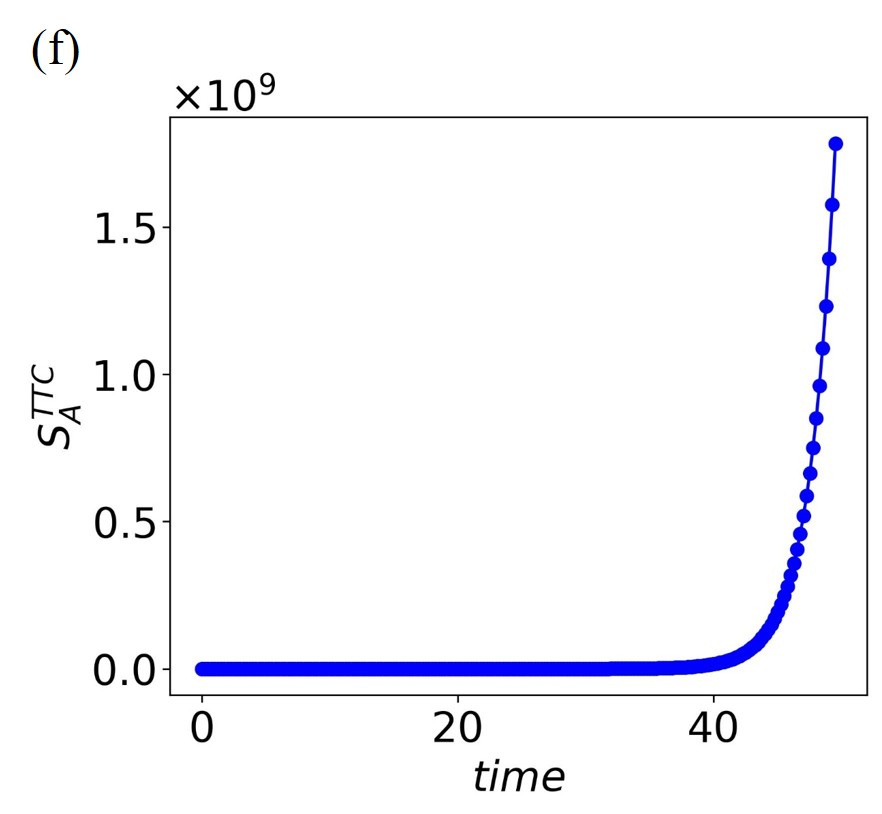}
\end{subfigure}
\begin{subfigure}[b]{0.4\linewidth}        
     \includegraphics[width=\linewidth]{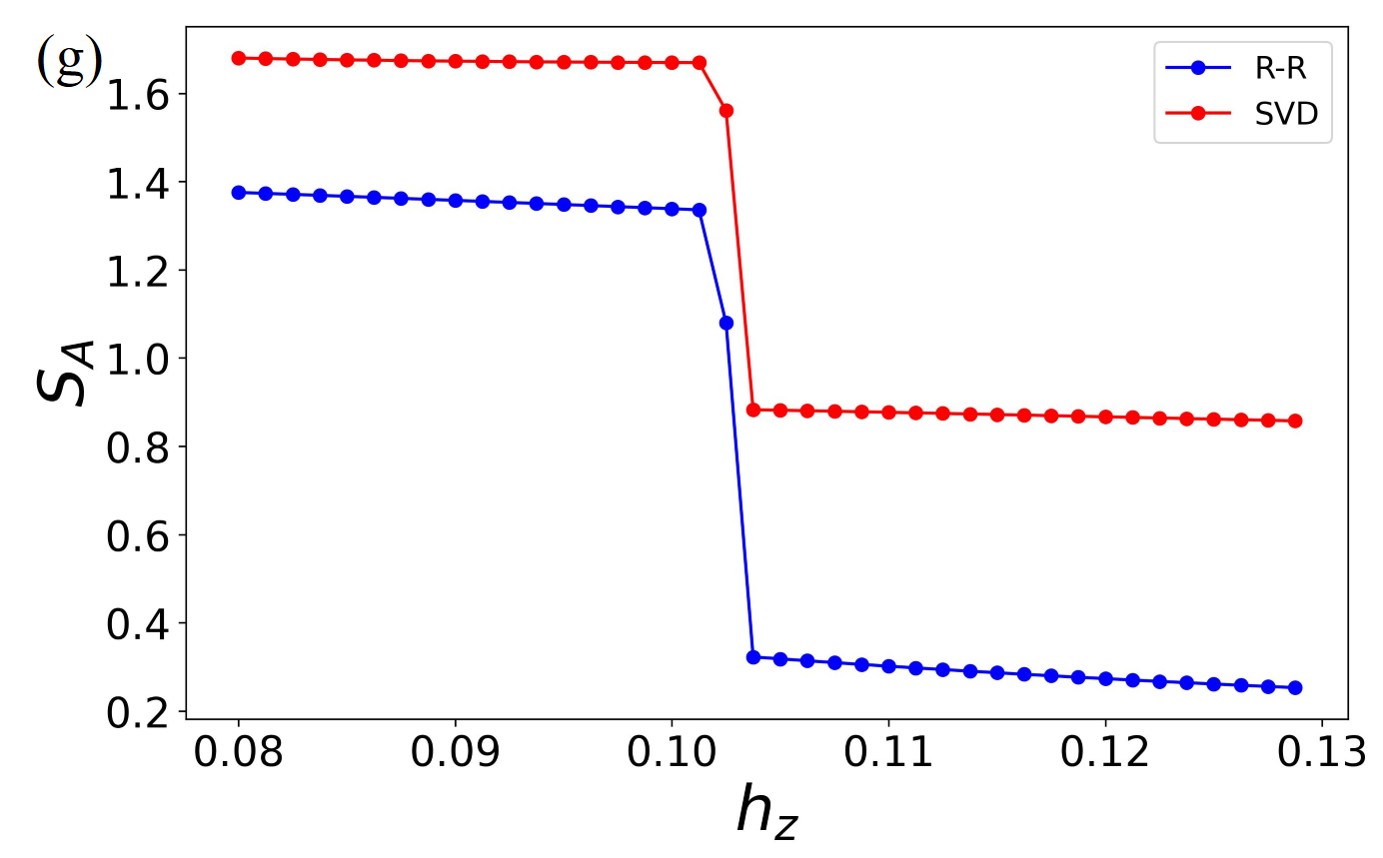}
\end{subfigure}
\caption{Entanglement dynamics in the Yang-Lee model for (a)(d) entanglement entropy, (b)(e) SVD entropy, and (c)(f) TTC entropy. 
 Upper panel (a-c) is for the $\mathcal{PT}$-unbroken case where the parameters are  $(J_1, J_2,h_x, h_z)=(0.4, 0.1, 0.9,0.03)$. 
  All three entropy measures exhibit unitary-like oscillations~\cite{PhysRevLett.128.010402}.
  Middle panel (e-f) is for the $\mathcal{PT}$-broken case where the parameters are  $(J_1, J_2,h_x, h_z)=(0.4, 0.1, 0.9,0.12)$.
  The entanglement entropy and the SVD entropy saturate to constant values at long time. The TTC entropy grows exponentially in time with rate $2E_I$ set by the maximum imaginary part of the spectrum.
 (g) The saturation value of the entanglement entropy and the SVD entropy have a volume to a area law transition crossing the level-crossing point, reflecting a change in the dominant eigenmode.}
  \label{fig:Yang_Lee_entropy}
\end{figure*}

For the $\mathcal{PT}$-broken case, at long time, the SVD density matrix can be expanded as follows
\begin{align}
    (\rho_A^{RL})^\dagger \rho_A^{RL}  &\sim P_0^\dagger P_0 + e^{E_It} (P_1^\dagger P_0 + P_0^\dagger P_1) + e^{2E_It} (P_1^\dagger P_1)\notag \\
    & + e^{3E_It} (P_2^\dagger P_1 + P_1^\dagger P_2) + e^{4E_It}(P_2^\dagger P_2) \notag \\
    &\sim e^{4E_It}(P_2^\dagger P_2).
\end{align}
At long time, the normalized SVD-RDM is
\begin{align}
\rho_A^{1|2} \xrightarrow{t \to \infty}  \frac{\sqrt{(P^A_2)^\dagger P^A_2}}{ {\rm Tr} \sqrt{(P^A_2)^\dagger P^A_2} } \propto {\rm Tr }_B \sqrt{| L_{\rm min}  \rangle \langle  L_{\rm min}  |}.
\end{align}
At long time, the SVD entropy saturates to a constant. 
Compare to the $\rho^{RR}$, we will find that the dominated terms in not the maximum imaginary parts of energy spectrum, instead the minimum imaginary parts in left basis.
Both $|R_{\rm max}\rangle$ and  $|L_{\rm min}\rangle$ are originated from the ground and first excited state with a low entanglement property.
Similar to the entanglement entropy, the SVD entanglement entropy will under go to a volume law to a area-law transition crossing the level-crossing point [see Fig.~\ref{fig:Yang_Lee_entropy}(g)].

\section{Distinct quench dynamics of the TTC entropy between interacting and free-fermion systems}
\label{Free_interacting}
In generic interacting systems, as we discussed in the previous section, the TTC entropy grows exponentially when the system is under a sudden quench of the $\mathcal{PT}$-broken Hamiltonian.
However, for free-fermion systems under a sudden quench of the $\mathcal{PT}$-broken Hamiltonian, the TTC entropy decays linearly.
This linear decay can be understood from the property of the correlation matrix, where the TTC entropy can be computed directly from the correlation matrix in free-fermion systems.

\subsection{Linear decay of the TTC entropy in the \texorpdfstring{$\mathcal{PT}$}{PT}-broken free-fermion case}
Let us start at analyzing the dynamics of the two-point correlation function in free-fermion systems.
The reduced density matrix of a free fermionic system has a Gaussian form~\cite{Chang_2020,Peschel_2003,Peschel_2009}:

\begin{align}
    \rho_A = \frac{1}{Z} \exp\left(-\sum\limits_{\alpha, \beta} \mathcal{H}^E_{\alpha\beta} \phi_\alpha^\dagger \phi_\beta \right),
\end{align}
where $\phi_\alpha^{(\dagger)}$ is the fermionic (creation) operator and $\alpha$ is the label of the combination of position, spin, and orbital degrees of freedom.

Due to this Gaussianity, all the higher-order correlation functions can be captured by the two-point correlation function.
For non-Hermitian free-fermion systems, the two-point correlation function is
\begin{align}
    C_{ij}
    = \langle \psi_L | \hat{\phi}_i^\dagger \hat{\phi}_j | \psi_R \rangle.
\end{align}

The dynamics of the correlation function $C_{ij}(t)= \langle \psi_L (t) | \hat{\phi}_i^\dagger \hat{\phi}_j | \psi_R(t) \rangle$ is similar to that of a typical observable quantity $\langle \hat{O}(t) \rangle$,
which is oscillatory when the post-quench Hamiltonian is $\mathcal{PT}$-unbroken and exhibits exponential growth when the post-quench Hamiltonian is $\mathcal{PT}$-broken.
I.e., $C_{ij}(t) \propto e^{2 E_I t}$.

The TTC entropy can be directly obtained from $C_{ij}(t)$~\cite{Chang_2020,Chen_2024,Guo_2021}:
\begin{align}
    S^{TTC}_A (t)= -\sum_{\delta} \left[ \nu_{\delta}(t) \ln \big|\nu_{\delta}(t)\big| + (1 - \nu_{\delta}(t)) \ln \big|1 - \nu_{\delta}(t)\big| \right],
   \label{Eq:STT_free}
\end{align}
where \( \nu_{\delta}(t) \) are the eigenvalues of the correlation matrix \( C_{ij}(t) \). 
For the $\mathcal{PT}$-broken case, at late time $C_{ij}(t) \propto e^{2 E_I t}$, the TTC entropy has the form 
\begin{align}
    S^{\rm TTC}_A(t)
    &\sim - (e^{2 E_I t} \nu_{\delta0'}) \ln{ \bigg| \frac{e^{2 E_I t} \nu_{\delta0'}}{e^{2 E_I t} \nu_{\delta0'} - 1}  \bigg|} 
    - \ln{ |e^{2 E_I t} \nu_{\delta0'} - 1|} \notag \\
    &\xrightarrow{t \to \infty} - 2 E_I t.
\end{align}
where \( \nu_{\delta'} (t) = e^{2E_It} \nu_{\delta0'} \) is one of the eigenvalues of correlation matrix \( C_{ij}(t) \) with the largest exponential amplification.
This property is different from Eq.~(\ref{Eq:STT_int}) for generic (interacting) cases.
In free-fermion systems, the biorthogonal reduced density matrix has the form~\cite{Chang_2020,Peschel_2003}
\begin{align}
\rho^{RL}_A (t )= \bigotimes_\delta  \left(\begin{array}{cc}\nu_\delta (t) & 0 \\0 & 1- \nu_\delta (t)\end{array}\right).
\end{align}
At long time, the dominant eigenmode of the correlation matrix, $\nu_\delta (t) = e^{2E_It} \nu_{\delta0} $ and $1-\nu_\delta (t) = 1-e^{2E_It} \nu_{\delta0} $ leads to an approximate spectral symmetry of the biorthogonal RDM.
In other words, eigenvalues of the biorthogonal RDM form a set of $\{ \alpha_i e^{2E_It}, - \alpha_i e^{2E_It} + \epsilon \}$ with $\epsilon $ being some non-vanishing numbers.
At long time, the eigenvalues of the biorthogonal RDM come in approximately plus-minus pairs. This indicates that the spectrum of the biorthogonal RDM possesses an approximate $\pi$ rotation symmetry with respect to the origin of the complex plane. This spectral property of the biorthogonal RDM is generically broken when interaction is introduced.

We examine the distinct property of the TTC entropy under a sudden quench of a $\mathcal{PT}$-broken Hamiltonian, comparing the free-fermion and interacting cases free and interacting fermion cases from the non-Hermitian spin-$1/2$ XXZ chain~\cite{Tu2023}:
\begin{align}
    H 
    &= \sum_{j=1}^{L} \left( \sigma_j^x \sigma_{j+1}^x + \sigma_j^y \sigma_{j+1}^y + J_z \sigma_j^z \sigma_{j+1}^z \right)
    \notag \\
    &\quad + \sum_{j=1}^{L/2} i\gamma_j \left( \sigma_{2j-1}^z - \sigma_{2j}^z \right).
\end{align}
When \( J_z = 0 \), this model reduces to a free-fermion chain~\cite{Tu2023}. 
The \( J_z \) term introduces interaction between fermions and allows us to explore how interaction modifies quench dynamics of the TTC entropy.

\begin{figure}[htbp]
    \centering
    \begin{subfigure}[t]{0.49\linewidth}
        \centering
        \adjustbox{valign=t}{\includegraphics[width=\linewidth]{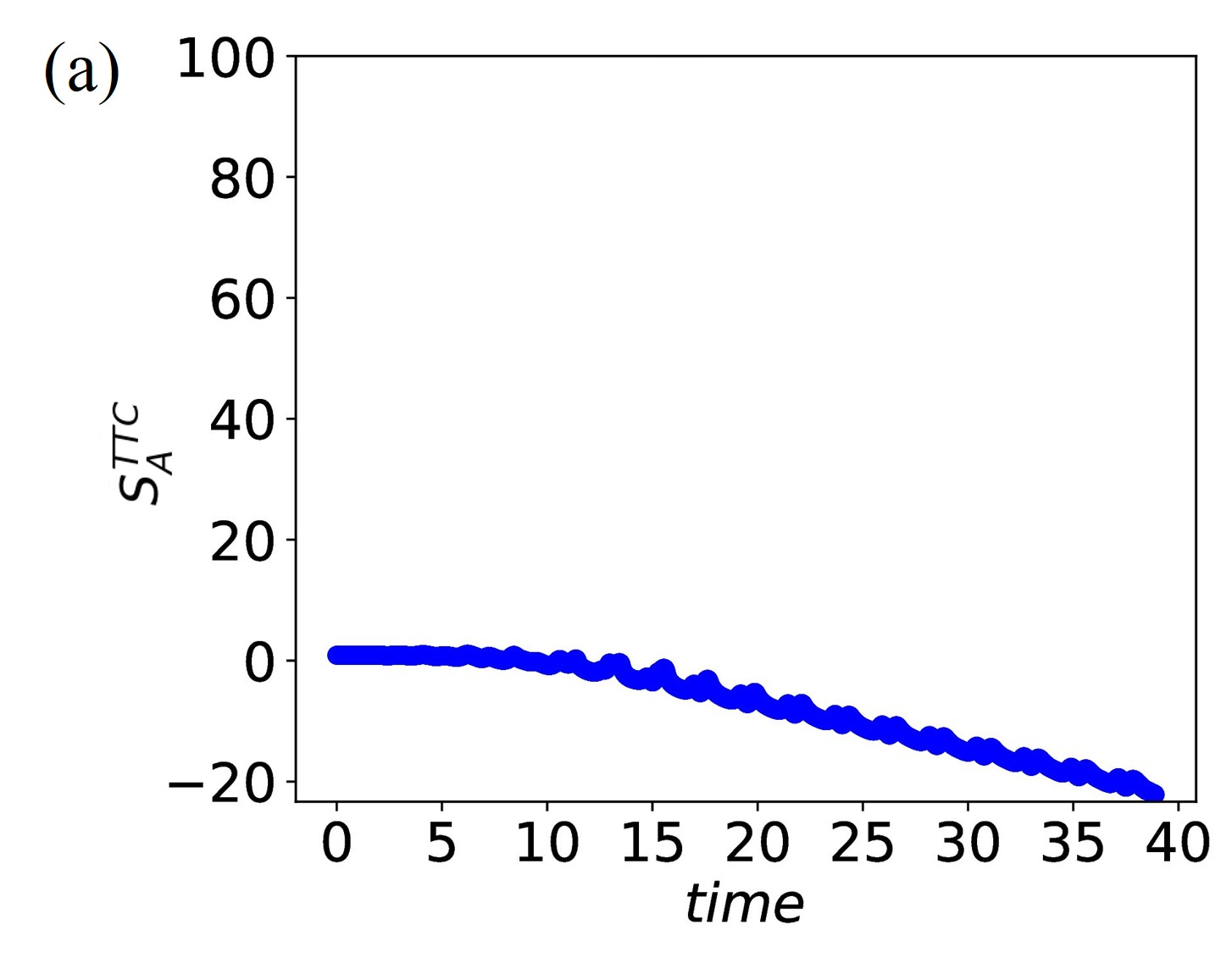}}
    \end{subfigure}
    \begin{subfigure}[t]{0.49\linewidth}
        \centering
        \adjustbox{valign=t}{\includegraphics[width=\linewidth]{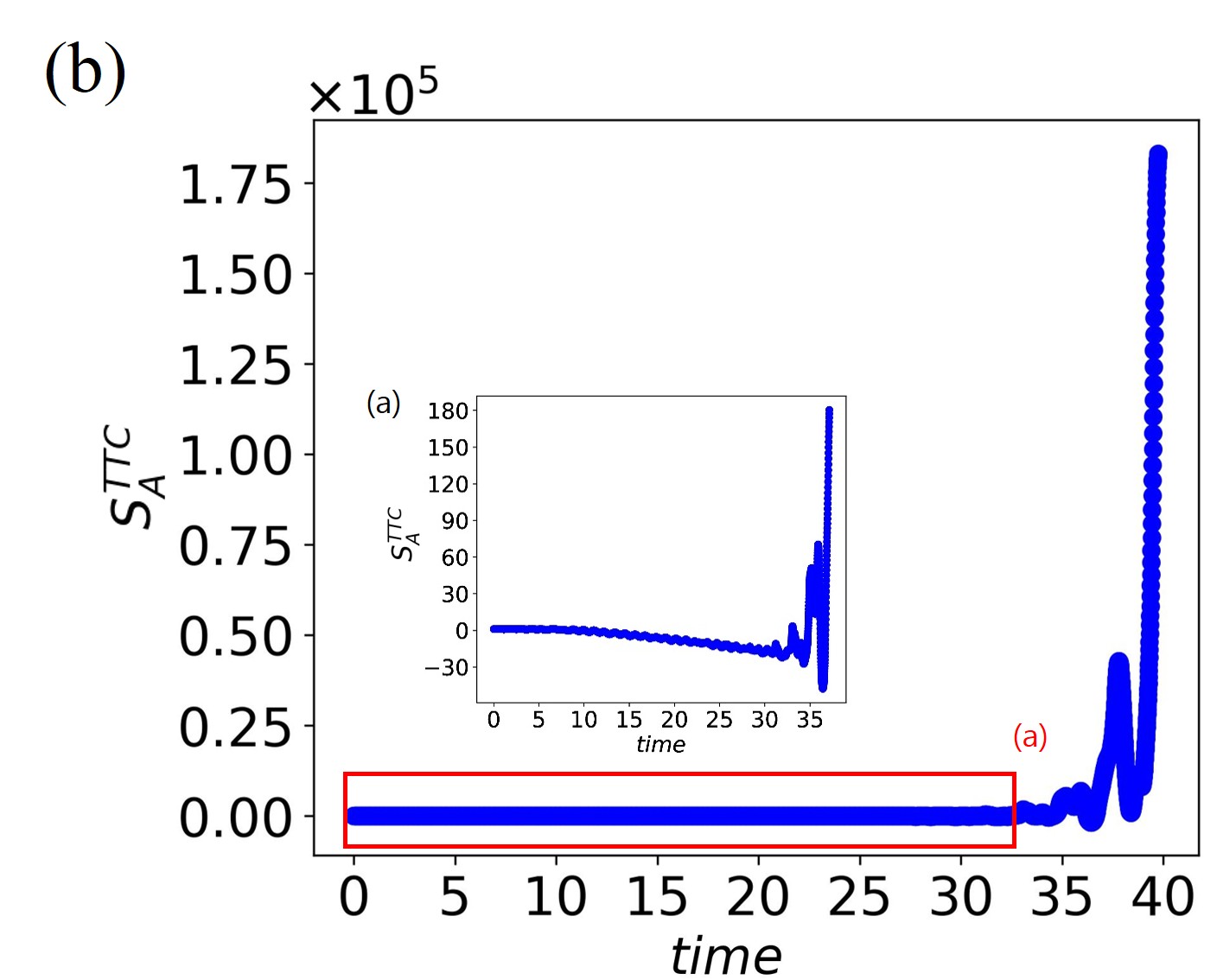}}
    \end{subfigure}
    \hfill
    \begin{subfigure}[t]{0.8\linewidth}
        \centering
        \adjustbox{valign=t}{\includegraphics[width=\linewidth]{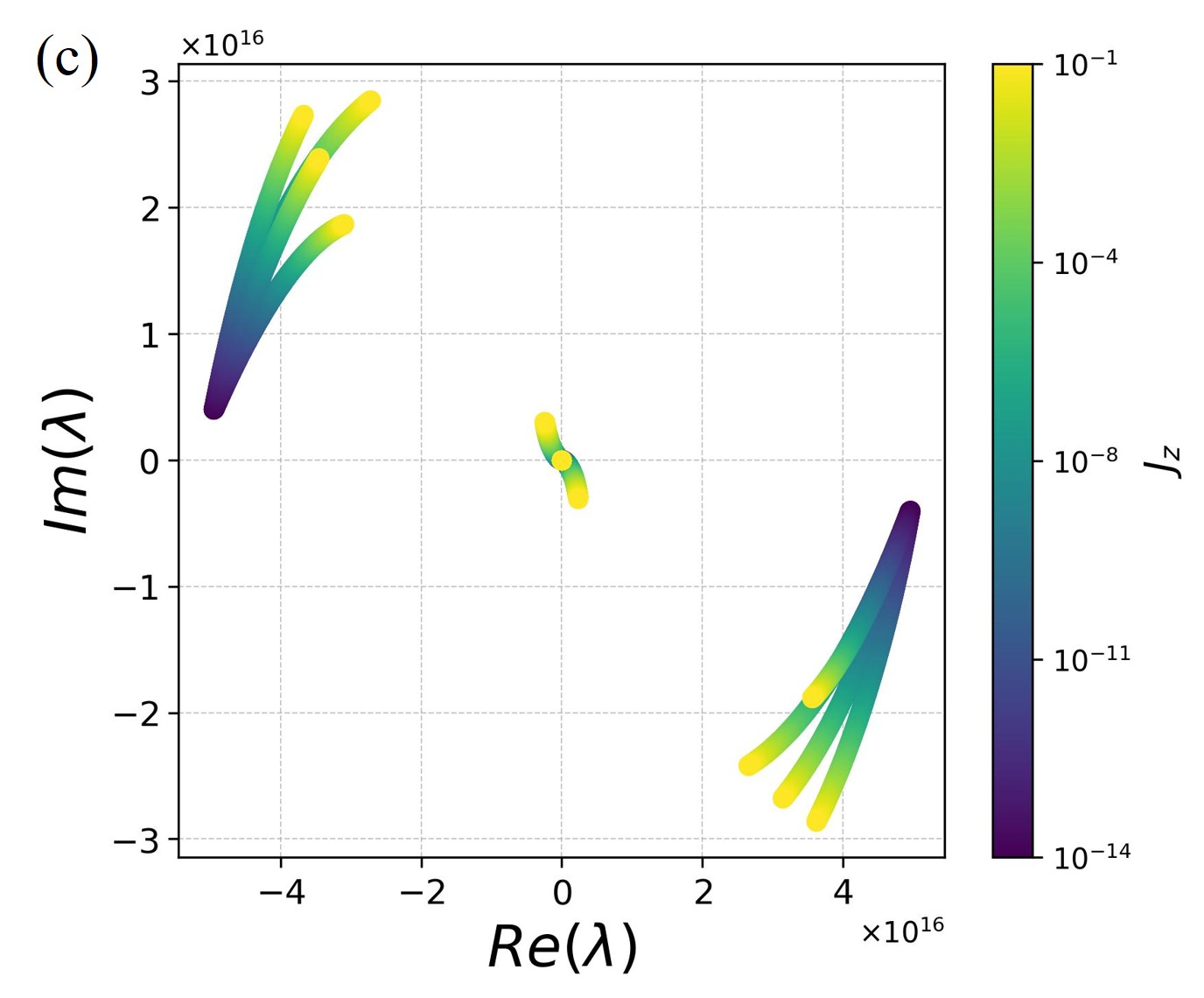}}
    \end{subfigure}
    \caption{Interaction effects of the TTC entropy in the non-Hermitian XXZ chain where the paramter is $\bar{\gamma_j}=0.1$. 
  (a) For $J_z=0$ (free-fermion case), the entanglement entropy decays linearly. 
   (b) With small interaction ($J_z=10^{-5}$), the entanglement entropy crosses over from linear to exponential growth. 
   (c) Eigenvalue trajectories of the biorthonormal RDM as function of $J_z$ at late time ($t=50$ and $1/2E_I \sim 2$). 
        The color map runs from $0$ (deep purple) to $J_z=10^{-4}$ (bright yellow). For very weak interaction, the eigenvalues 
   of the biothonormal RDM remain in approximate $\pm\lambda$ pairs (purple), giving rise to a linear TTC entropy decay. 
   When $J_z$ becomes large, the approximate $\pm\lambda$ pairs are destroyed (yellow), leading to exponential growth of the TTC entropy.}
    \label{fig:interaction_effects}
\end{figure}

As shown in Fig.~\ref{fig:interaction_effects}(a), in the non-interacting case ($J_z = 0 $), the TTC entropy decays linearly in time $S^{\rm TTC}_A (t) \sim -2 E_I t$.
Once interaction is introduced ($J_z \neq 0$), the exponential growth of the  TTC entropy is observed at late time [see Fig.~\ref{fig:interaction_effects}(b)].
The spectrum of the biorthogonal RDM at late-time as a function of the interaction strength $J_z$ is shown in Fig.~\ref{fig:interaction_effects}(c).
The spectrum is approximately symmetric for $J_z=0$. For small but nonvanishing $J_z$, the approximate spectral symmetry is broken as indicated by light-green/yellow dots in Fig.~\ref{fig:interaction_effects}(c).



\section{Conclusion}
\label{Conclusion}
In this work, we investigated the quench dynamics of observable quantities, the quantum geometric tensor, and various entanglement entropies in $\mathcal{PT}$-symmetric non-Hermitian systems using the biorthogonal basis. 
Although the trace-preserving property of the biorthogonal density matrix holds ${\rm Tr}  \rho^{RL}(t)=1$, the non-positivity of the biorthogonal density matrix leads to exponential growth of all quantities we studied for the $\mathcal{PT}$-broken case.
One interesting observation is that the exponential growth of the TTC entropy becomes linear when the system is non-interacting.
The origin of this linear-growth TTC entropy comes from the approximate spectral symmetry of the biorthogonal reduced density matrix.
We demonstrate these properties for the Yang-Lee model and the non-Hermitian spin-1/2 XXZ chain.
Experimentally, physical quantities such as the biorthogonal chiral displacement have been measured through the biorthogonal formalism~\cite{Wang2021}.
One interesting direction is to experimentally implement the biorthogonal quench dynamics in the $\mathcal{PT}$-broken case.

\begin{acknowledgments}
P.-Y. C. thanks Xiao Chen for insightful discussions.
We acknowledge support from
the National Science and Technology Council of Taiwan
under Grants No. NSTC 113-2112-M- 007-019, 114-2918-I-007-015, and the support from Yukawa Institute for Theoretical Physics,
Kyoto University,  RIKEN Center for Interdisciplinary Theoretical and Mathematical Sciences, and  National Center for Theoretical Sciences, Physics Division.
\end{acknowledgments}

\appendix
\section{Comparison of Maximum Imaginary Part of Energy Spectrum}

In this appendix, we investigate how different parameters of the Yang-Lee model influence the maximum imaginary part of the energy spectrum and the resulting dynamical behavior. In particular, we focus on the effects of the next-nearest neighbor coupling $J_2$ and the transverse field $h_x$, while keeping the nearest-neighbor coupling $J_1=0.4$ (as in the main text). By analyzing the spectrum for various $J_2$ (e.g.\, $0,\,0.001,\,0.1$) and different $h_x$ values, we can determine how the onset of complex eigenvalues ($\mathcal{PT}$-symmetry breaking) and the occurrence of level crossings in the energy levels are affected. 

\begin{figure}[t]  
\begin{subfigure}[b]{0.3\linewidth}        
     \includegraphics[width=\linewidth]{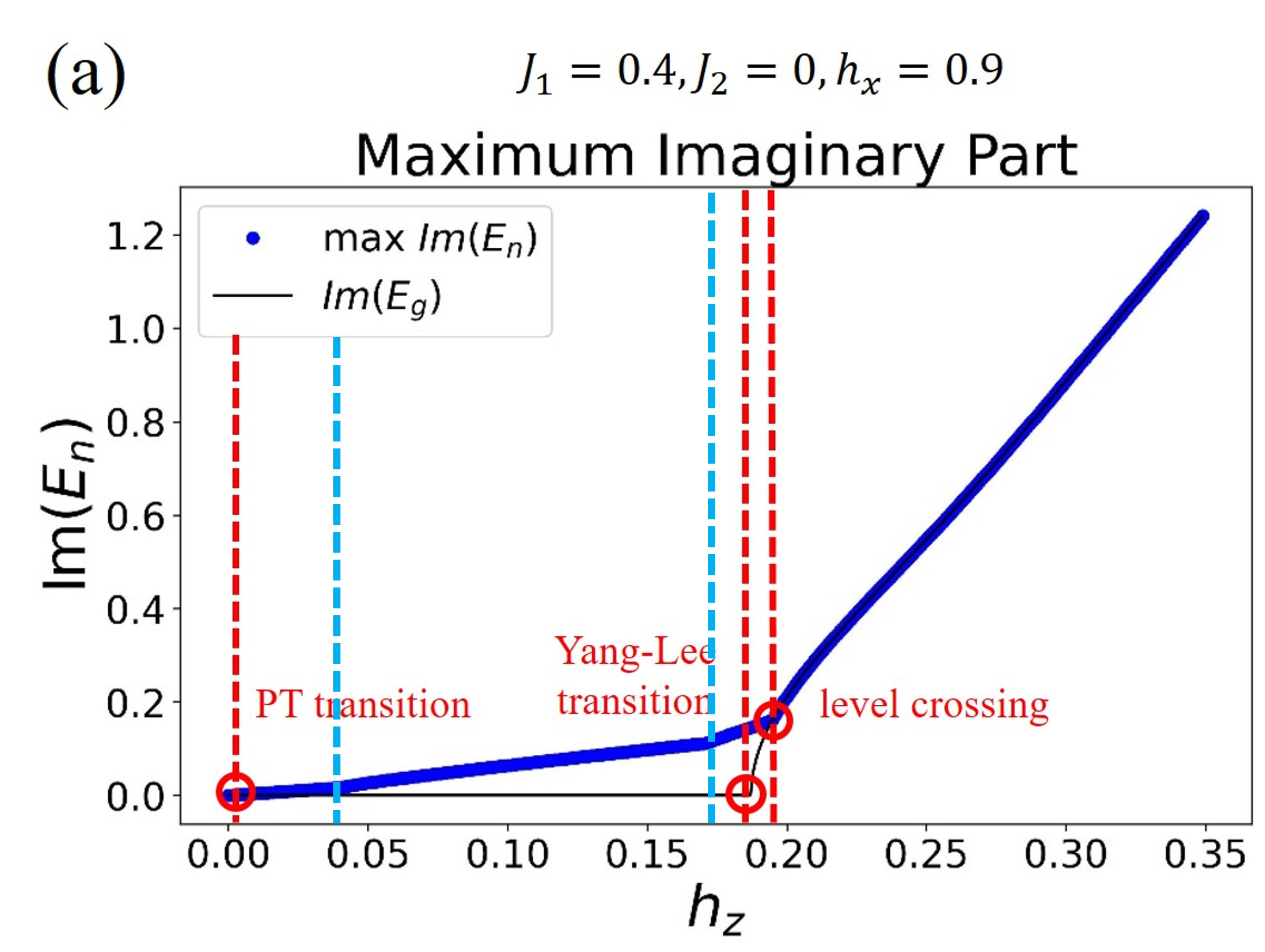}
\end{subfigure}
\begin{subfigure}[b]{0.3\linewidth}        
     \includegraphics[width=\linewidth]{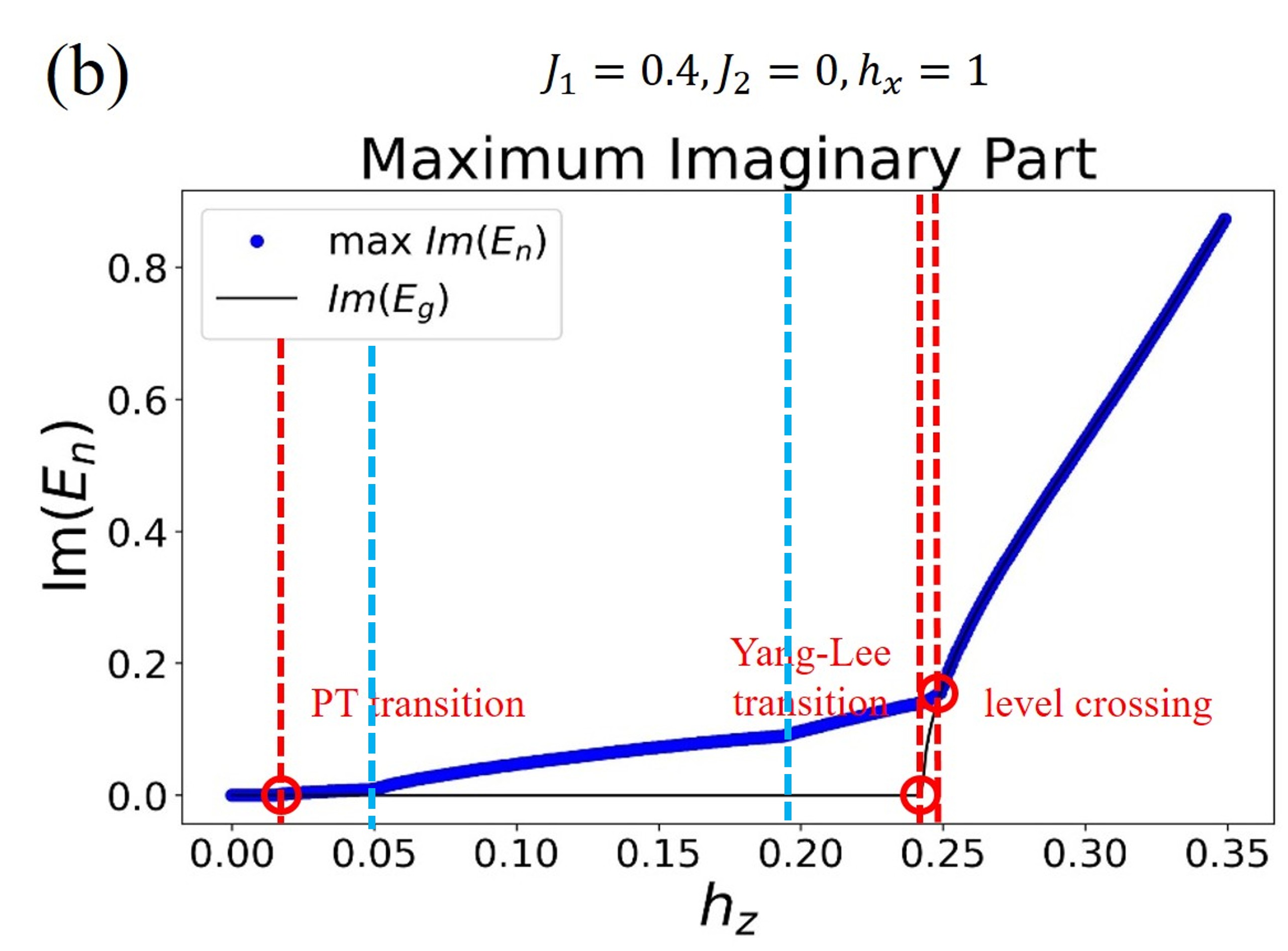}
\end{subfigure}
\begin{subfigure}[b]{0.3\linewidth}        
     \includegraphics[width=\linewidth]{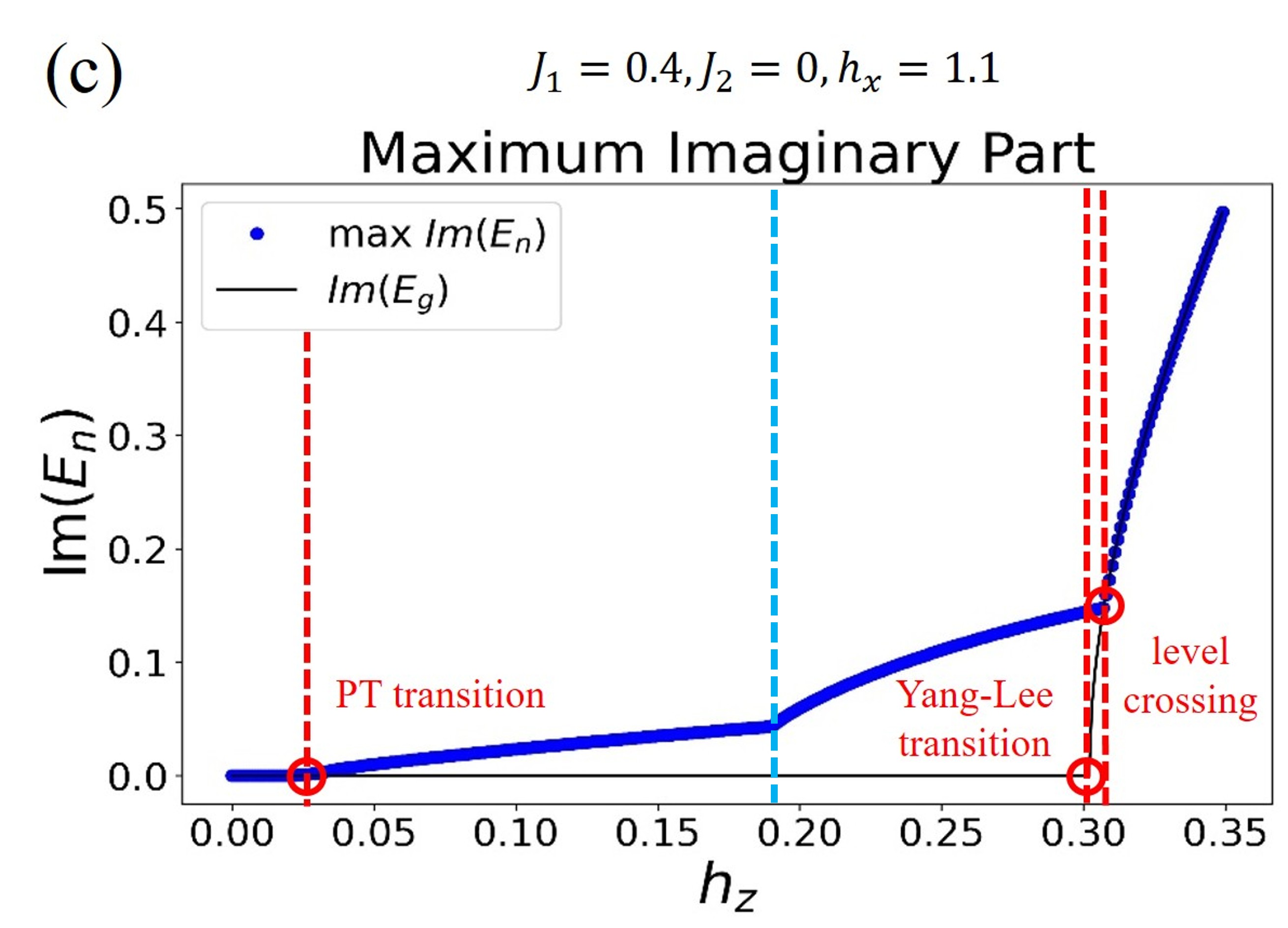}
\end{subfigure}
\begin{subfigure}[b]{0.3\linewidth}        
     \includegraphics[width=\linewidth]{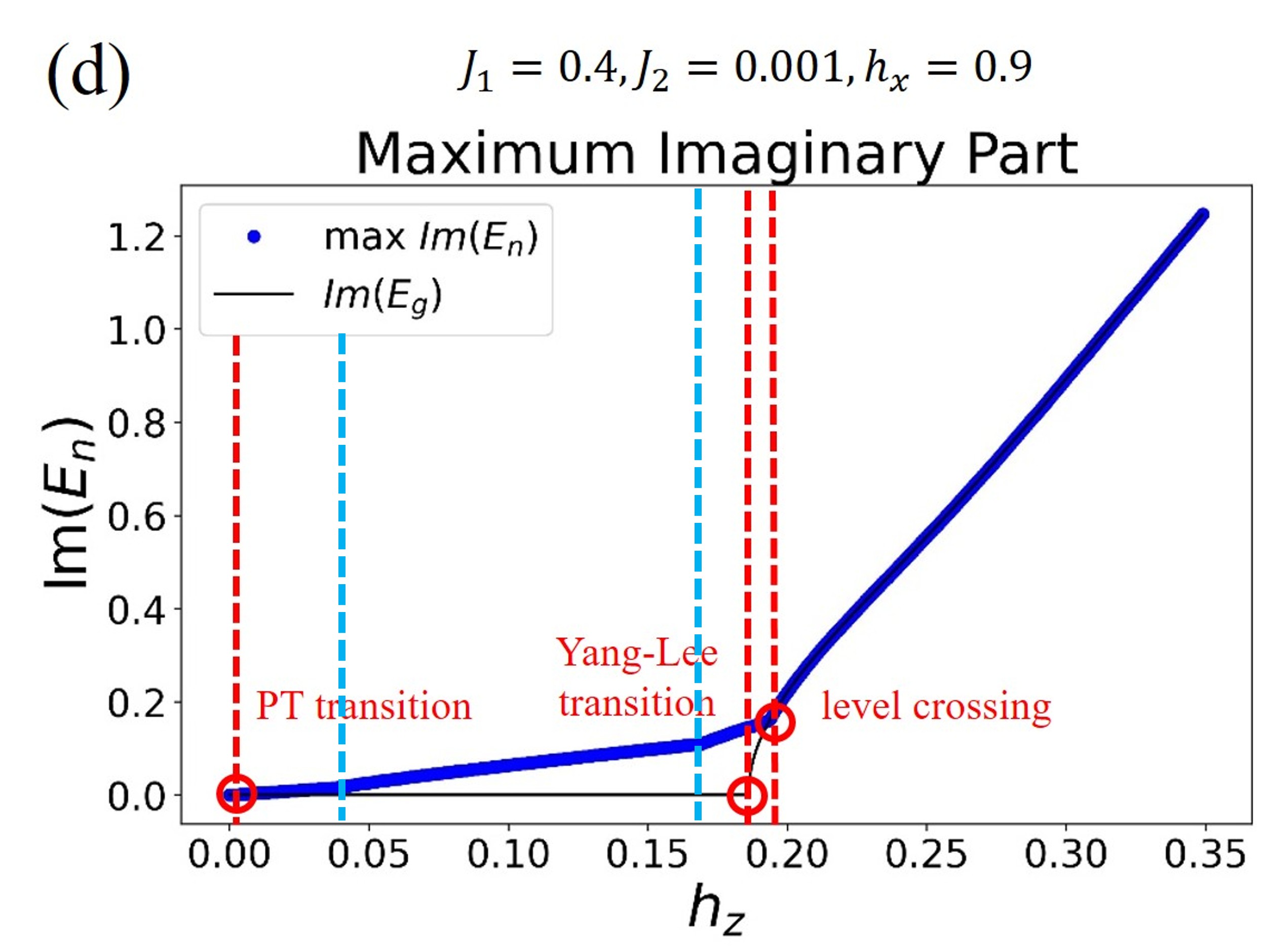}
\end{subfigure}
\begin{subfigure}[b]{0.3\linewidth}        
     \includegraphics[width=\linewidth]{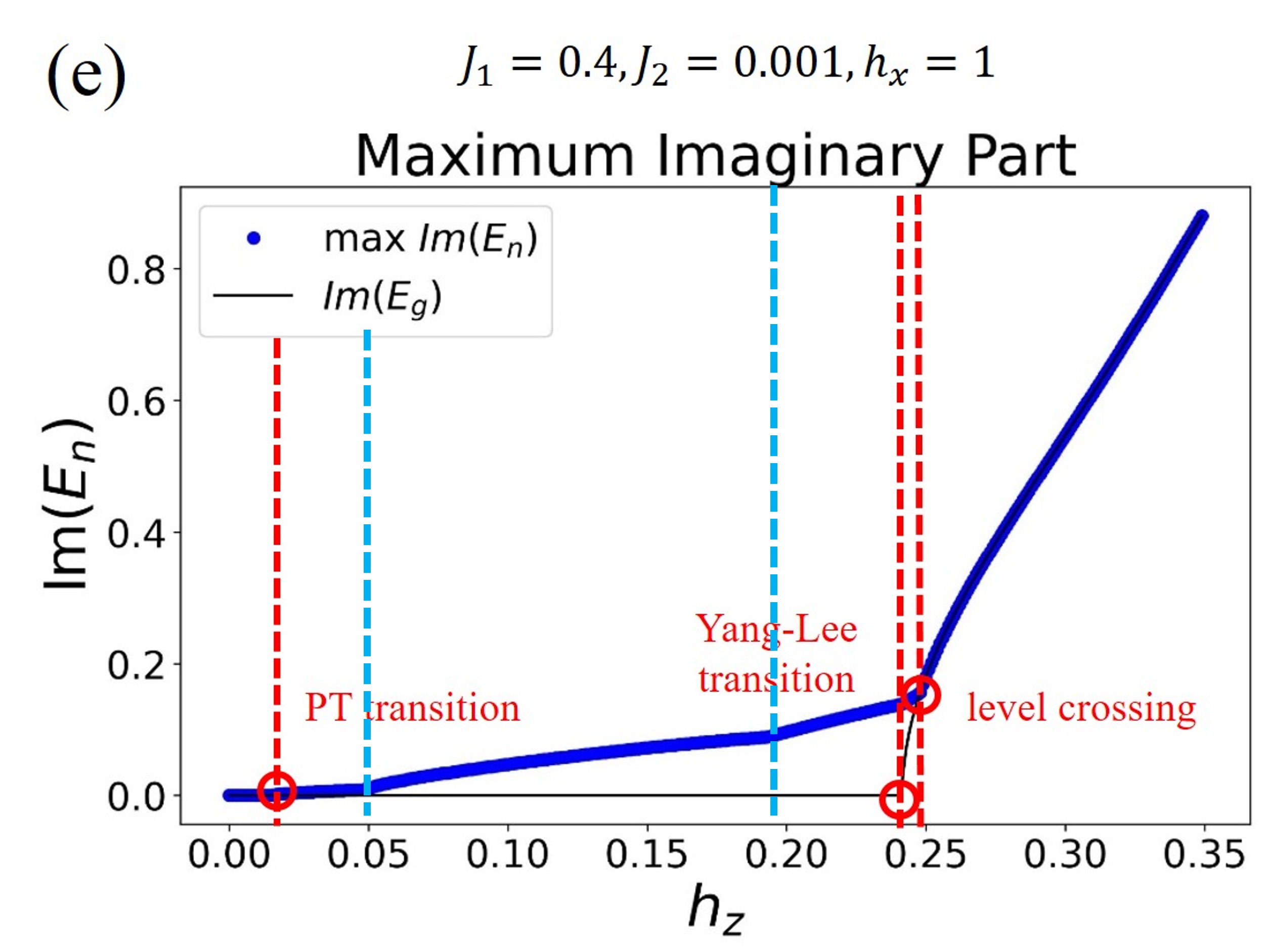}
\end{subfigure}
\begin{subfigure}[b]{0.3\linewidth}        
     \includegraphics[width=\linewidth]{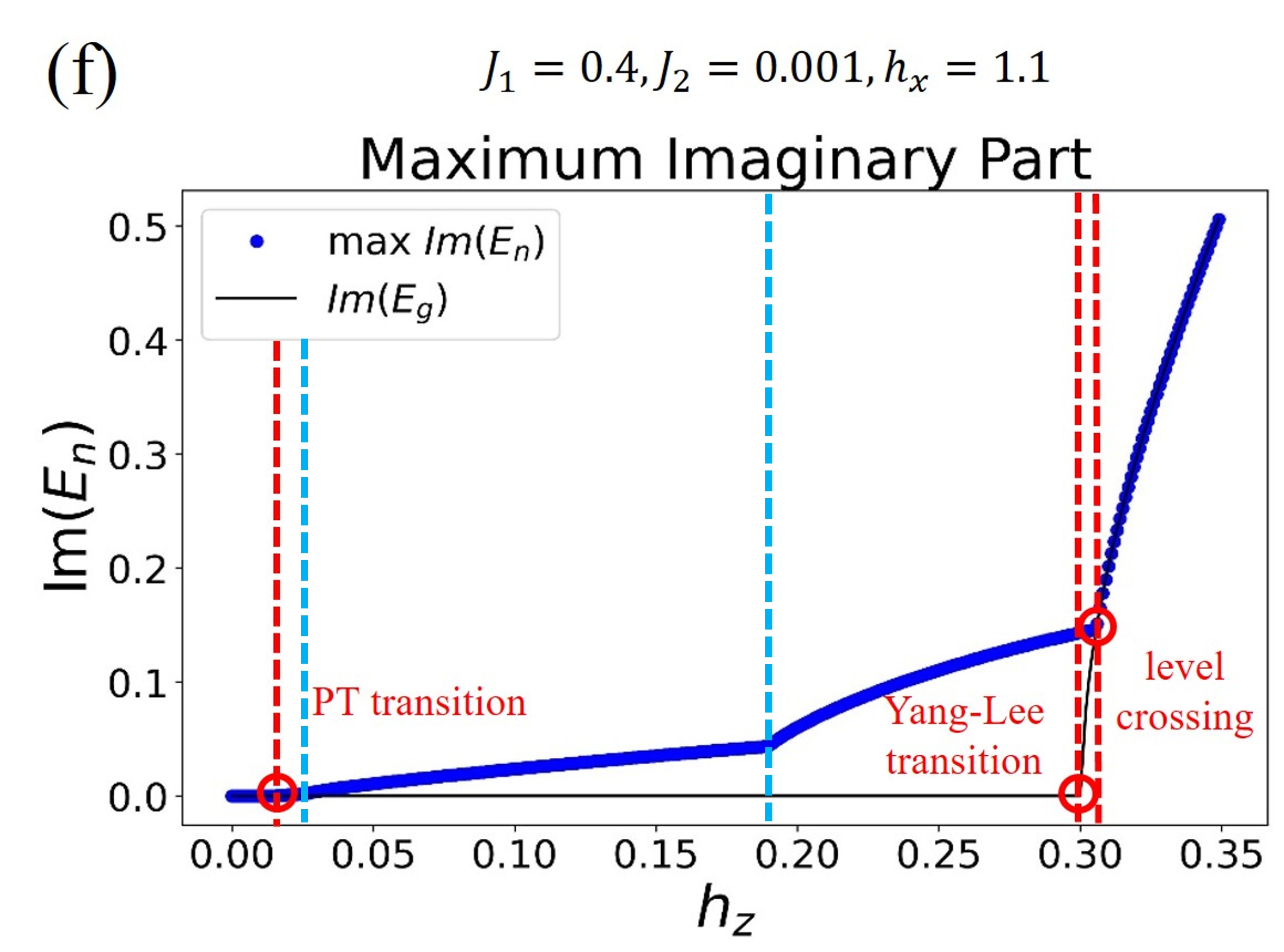}
\end{subfigure}
\begin{subfigure}[b]{0.3\linewidth}        
     \includegraphics[width=\linewidth]{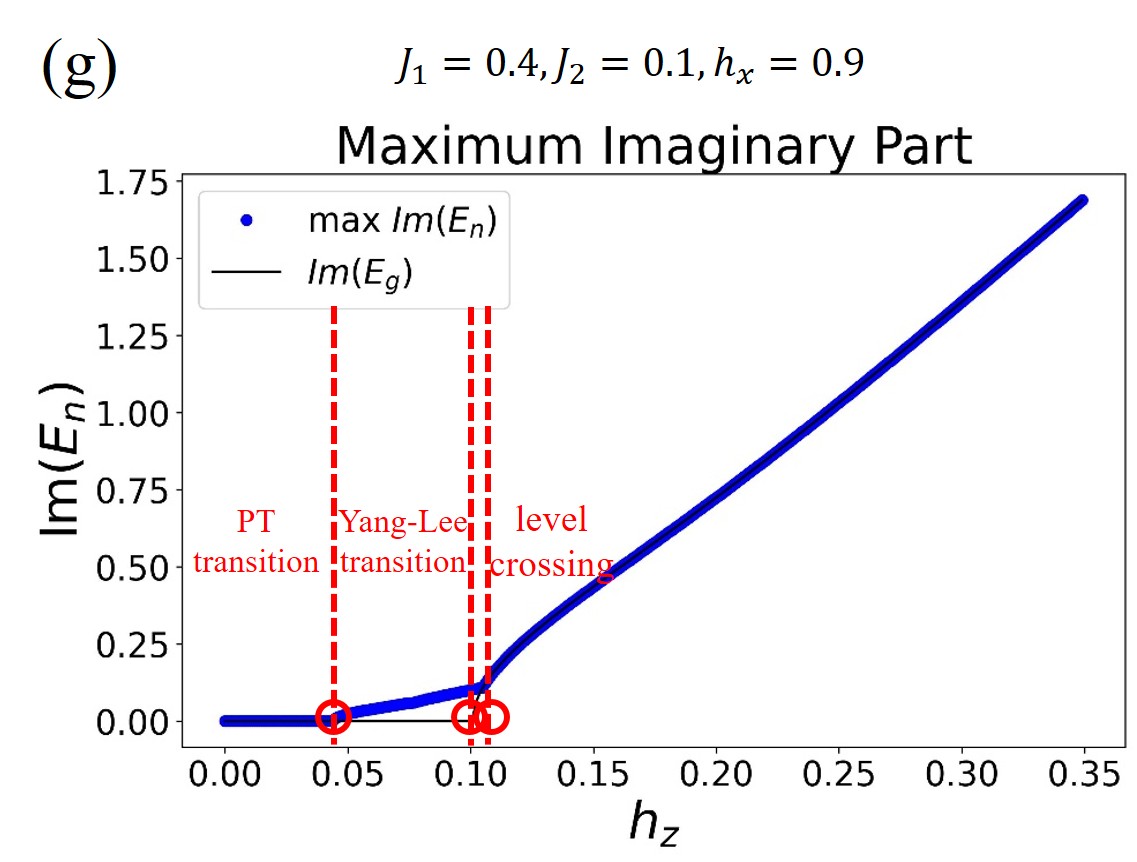}
\end{subfigure}
\begin{subfigure}[b]{0.3\linewidth}        
     \includegraphics[width=\linewidth]{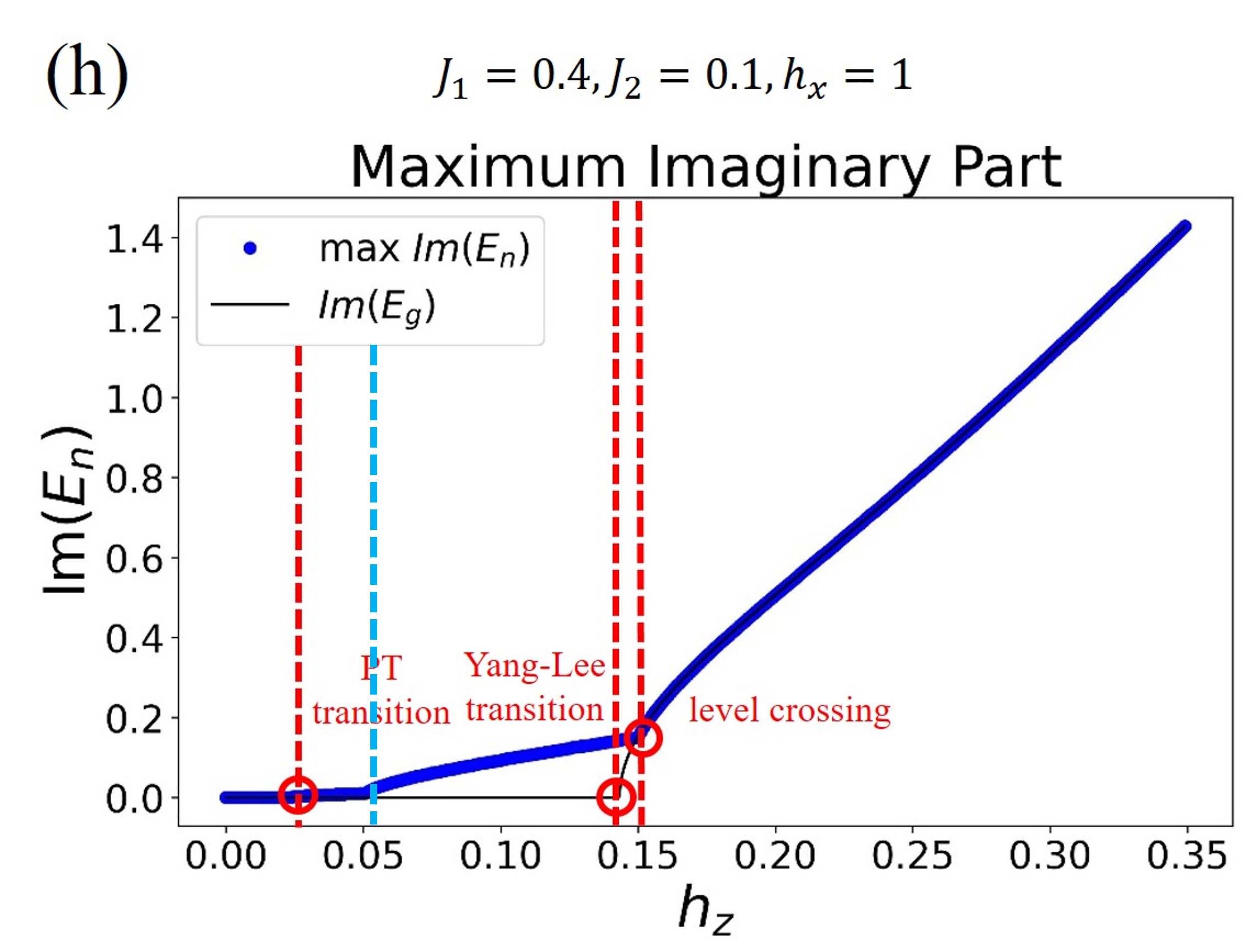}
\end{subfigure}
\begin{subfigure}[b]{0.3\linewidth}        
     \includegraphics[width=\linewidth]{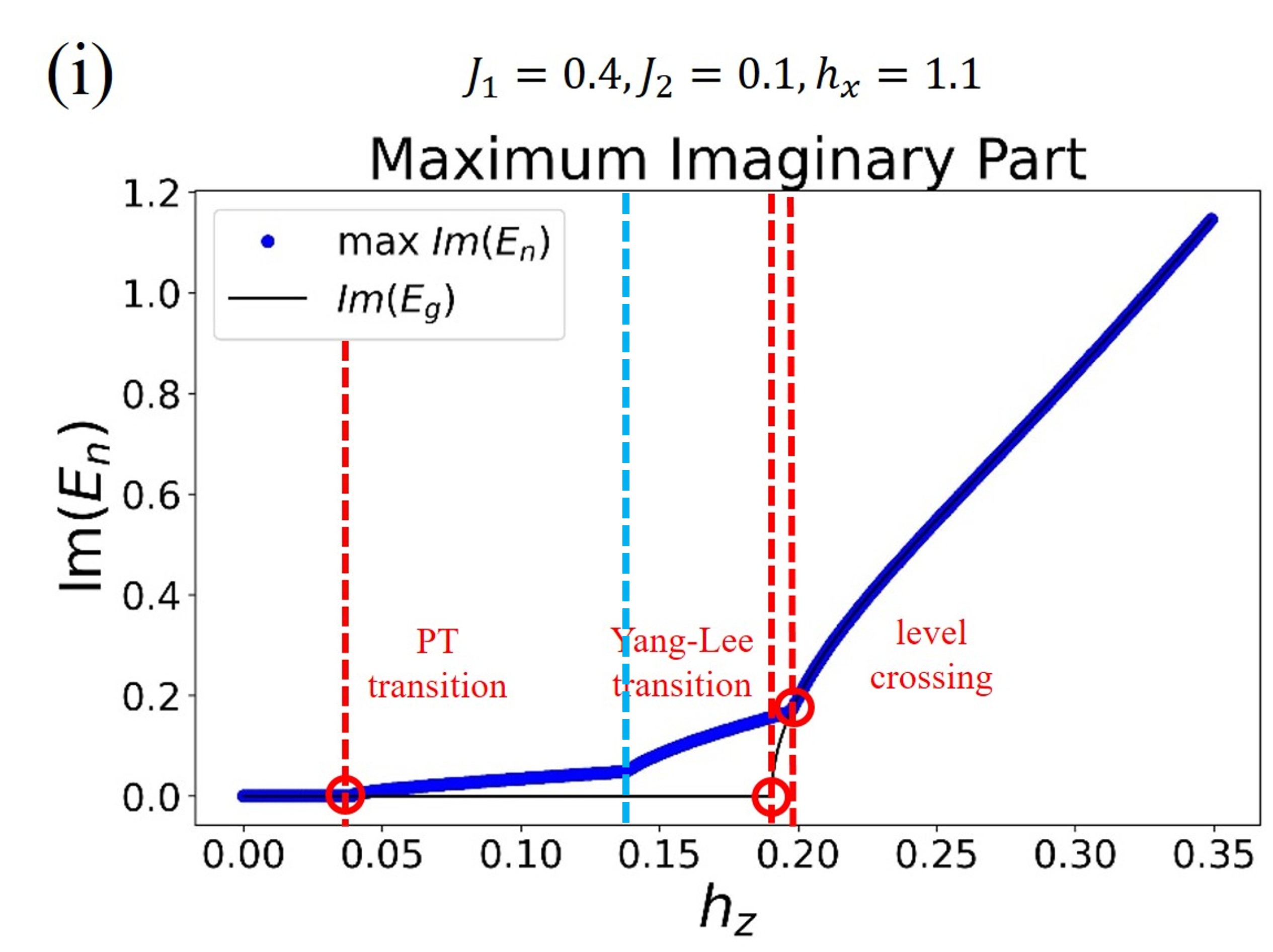}
\end{subfigure}
\caption{Maximum
imaginary part of the energy spectrum in the Yang-Lee model for different transverse field (a)(d)(g) $h_x=0.9$, (b)(e)(h) $h_x=1$, and
(c)(f)(i) $h_x=1.1$. The blue dashed line denotes a level crossing with another state (not the ground state). (a-c) is for no NNN coupling $J_2=0$; (d-f) is for a very small NNN coupling $J_2=0.001$; (g-i) is for a relatively large NNN coupling $J_2=0.1$.}
  \label{fig:app-A}
\end{figure}

We now compare three representative cases for the next-nearest neighbor (NNN) coupling: $J_2=0$ (no NNN coupling), $J_2=0.001$ (a very small NNN coupling), and $J_2=0.1$ (a relatively large NNN coupling). Figure~\ref{fig:app-A} summarizes the results for the maximum imaginary part of the energy spectrum as a function of the imaginary field $h_z$ for these different $J_2$ values (at given transverse field $h_x$ values). From these results, we observe that introducing a finite $J_2$ significantly shifts the $\mathcal{PT}$-breaking point. In the absence of NNN interactions ($J_2=0$), the critical $h_z$ at which the spectrum becomes complex is notably small – on the order of $10^{-3}$ (one order of magnitude smaller than in the coupled cases). As we turn on a nonzero $J_2$, the $\mathcal{PT}$-breaking threshold moves to larger values of $h_z$. For example, even a tiny coupling $J_2=0.001$ increases the critical $h_z$, and a stronger coupling $J_2=0.1$ pushes the onset of complex eigenvalues to a much higher $h_z$ (on the order of $10^{-2}$). In effect, the next-nearest neighbor coupling stabilizes the $\mathcal{PT}$-symmetric phase by allowing the system to maintain all-real eigenvalues up to a larger imaginary field. This means that the presence of $J_2$ delays the $\mathcal{PT}$-breaking point to a higher $h_z$ regime. 

In addition to shifting the $\mathcal{PT}$-breaking point, the next–nearest–neighbor coupling $J_2$ also affects the Yang–Lee edge singularity and subsequent level crossings in the energy spectrum, particularly those involving the ground state. A level crossing indicates that the ground state becomes the eigenstate with the maximum imaginary part, which then dominates the long‑time dynamics. In our model, we find that introducing $J_2$ causes both the Yang–Lee edge singularity and the corresponding ground‑state level crossing to occur at a lower value of $h_z$ (i.e., earlier) than in the $J_2=0$ case, for a fixed transverse field $h_x$. Moreover, without NNN coupling, the spectrum as a function of $h_z$ can exhibit multiple level crossings among $\mathcal{PT}$‑broken states after the $\mathcal{PT}$‑breaking point. In the $J_2=0$ scenario, as $h_z$ increases beyond its small critical value, several $\mathcal{PT}$‑broken levels may cross in succession, leading to multiple reconfigurations of which state dominates the quench dynamics. By contrast, when a substantial NNN coupling (e.g.\ $J_2=0.1$) is present, the spectrum is more widely spaced and only a single prominent level crossing involving the ground state occurs for $h_z$ up to the $\mathcal{PT}$‑breaking threshold. In other words, adding $J_2$ simplifies the evolution of the level structure by removing (or shifting beyond the region of interest) secondary crossings, leaving a single dominant ground‑state transition in the $\mathcal{PT}$‑broken phase.

For clarity in our analysis, we select model parameters that yield a single ground‑state level crossing within the regime of interest while also placing the $\mathcal{PT}$‑breaking threshold at a conveniently large value of $h_z$. Concretely, we employ a finite next‑nearest‑neighbor coupling and an appropriate transverse field. For instance, in Fig.~\ref{Fig:1} we use 
\(
(J_1, J_2, h_x) = (0.4,\;0.1,\;0.9),
\)
for which the $\mathcal{PT}$‑breaking point occurs at $h_z \approx 0.045$, and only one level crossing—between the ground state and the state with maximum imaginary part—appears up to that point. This choice shifts the $\mathcal{PT}$‑breaking threshold to the $10^{-2}$ scale, simplifying the comparison of pre‑ and post‑breaking dynamics and ensuring that there is no other level crossing between $\mathcal{PT}$-breaking point and Yang-Lee edge singularity. Such a setup allows us to clearly capture the sharp, first‑order jump in the system’s entanglement entropy induced by that singularity.  

\bibliography{apssamp}

\end{document}